\documentclass[aps,prl,preprintnumbers,twocolumn,nofootinbib]{revtex4-1}%,showpacs %nofootinbib %preprintnumbers, %superscriptaddress
\usepackage[dvips]{graphicx}
\usepackage{bm,latexsym,amsmath,amssymb,amsfonts,mathrsfs}
\usepackage{color}
\usepackage{braket}
\usepackage{latexsym}
\usepackage{wrapfig}
\usepackage{tabularx}
\usepackage{comment}
\usepackage{lipsum}
\usepackage{mathtools}

\input{colordvi.tex}
\newcommand{\coloneqq}{:=}
\newcommand{\der}{\partial}
\newcommand{\no}{\nonumber}
\newcommand{\re}{\mathrm{Re}}
\newcommand{\im}{\mathrm{Im}}

\newcommand{\pmass}{m_{\rm ph}}
\newcommand{\Mpl}{M_{\rm pl}}
\newcommand{\Ms}{M_{\rm s}}
\newcommand{\Lag}{\mathcal{L}}
\newcommand{\lag}{\mathcal{L}}
\newcommand{\lth}{\Lambda_{\rm th}}
\newcommand{\lcrit}{\Lambda_*}
\newcommand{\impr}{c_{2,\,{\rm impr}}}
\newcommand{\cgr}{c_\text{grav}}
\newcommand{\cgrt}{c_{{\rm grav},\,t\text{-}{\rm ch}}}

\newcommand{\cgro}{c_{{\rm grav,\,others}}}
\newcommand{\cngr}{c_\text{non-grav}}
\newcommand{\cngrp}{c_{{\rm non\text{-}grav,\,ren}}}
\newcommand{\scat}{\mathcal{M}}
\newcommand{\loopint}{\int\frac{\mathrm{d}^d\ell}{(2\pi)^d}}

\newcommand{\loopintt}{\int\frac{\mathrm{d}^dr}{(2\pi)^d}}
\newcommand{\MSb}{\overline{{\rm MS}}}

\begin{document}

%\preprint{KOBE-COSMO-21-07}

\title{Gravitational Positivity Bounds on Scalar Potentials}

\author{Toshifumi Noumi}
%\affiliation{Department of Physics, Kobe University, Kobe 657-8501, Japan}

\author{Junsei Tokuda}

\affiliation {Department of Physics, Kobe University, Kobe 657-8501, Japan}

\begin{abstract}
We derive constraints on scalar field theories coupled to gravity by using recently developed positivity bounds in the presence of gravity. It is found that a canonically-normalized real scalar cannot have an arbitrarily flat potential unless some new physics enters well below the Planck scale. An upper bound on the scale of new physics is determined by loop corrections to the self-energy. Our result provides a swampland condition for scalar potentials. 
\end{abstract}

\maketitle
%%%%%%%%%%%%%%%%%%%%%%%%%%%%%%%%%%%%%%%%%%%%%%%%%%%%%%%%%%%%%%%%%%%%%%

\section{I. Introduction} Scalar fields play an important role in various contexts of physics. In particle physics, the Higgs boson is a key ingredient of the Standard Model. In cosmology, we need the inflaton to realize the early universe inflation. More theoretically, moduli fields are crucial to understand the Landscape of quantum field theory models. In these contexts, it is important to clarify what kind of scalar potentials have a consistent ultraviolet (UV) completion, especially in the presence of gravity.

A starting point in this direction is the widely accepted statement that quantum gravity prohibits exact global symmetries and so completely flat potentials are not allowed~\cite{Banks:1988yz,Banks:2010zn,Harlow:2018tng}. Then, the question is how one can formulate more quantitative constraints  useful for phenomenology. Indeed, several bounds on scalar potentials have been proposed in the Swampland Program~\cite{Vafa:2005ui} with various degrees of rigors and motivations~\cite{Ooguri:2006in,ArkaniHamed:2006dz,Palti:2017elp,Obied:2018sgi,Garg:2018reu,Ooguri:2018wrx,Gonzalo:2019gjp,Benakli:2020pkm,Gonzalo:2020kke} (see also~\cite{Palti:2019pca, vanBeest:2021lhn} for reviews). The conjectured bounds, if true, have interesting implications for particle physics and cosmology, which motivates further studies toward their derivation.

In this paper, we explore quantum gravity constraints on scalar potentials in light of recently developed gravitational positivity bounds~\cite{Hamada:2018dde, Tokuda:2020mlf, Herrero-Valea:2020wxz}. In non-gravitational theories, it is well-known that Wilson coefficients of low-energy effective field theories (EFTs) have to satisfy an infinite set of inequalities called \emph{positivity bounds} in order to have a standard UV completion~\cite{Adams:2006sv}.
While its extension to gravitational theories has been non-trivial due to the graviton $t$-channel pole, the conditions under which (approximate) positivity bounds should hold are clarified by recent works~\cite{Hamada:2018dde, Tokuda:2020mlf, Herrero-Valea:2020wxz} (see \cite{Bellazzini:2019xts, Alberte:2020jsk, Caron-Huot:2021rmr} for related discussions).
Following this, we study a real scalar coupled to gravity in 4 dimensions,
\begin{equation}
\lag=\frac{\Mpl^2}{2}R-\frac{1}{2}(\der\phi)^2-V(\phi)
+\lag_{\rm higher}+\cdots
\,,\label{eq:action1}
\end{equation}
and use the gravitational positivity bounds to derive constraints on the scalar potential $V(\phi)$ and the higher derivative terms $\lag_{\rm higher}$, clarifying assumptions and limitation of its applicability. Here, $R$ and $\Mpl$ denote Ricci scalar and the reduced Planck mass, respectively. 

\medskip
\section{II. Gravitational positivity bounds}
\begin{figure}
\includegraphics[width=100mm, trim=200 220 150 200]{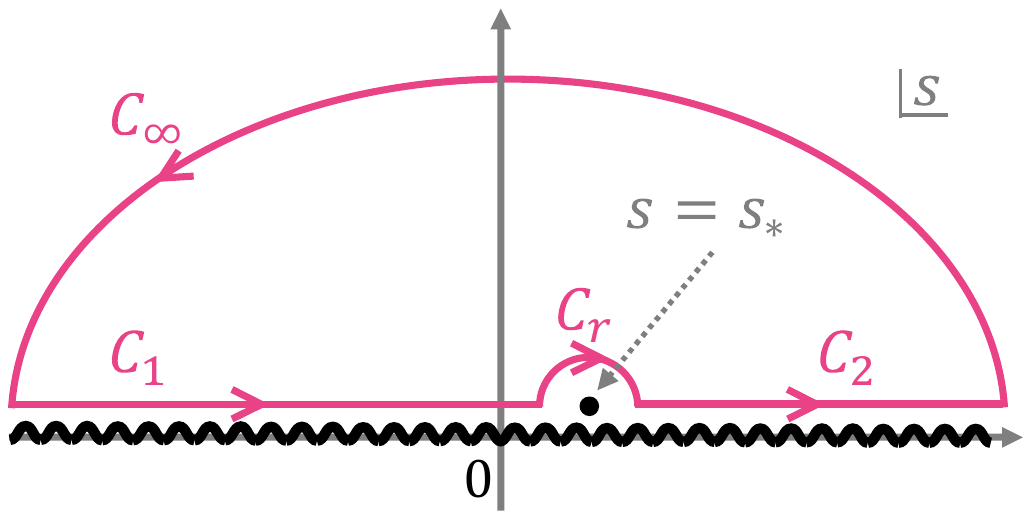}\caption{Analytic structure of $\widetilde\scat(s,t)$ on the complex $s$-plane and the integration contour to derive the relation \eqref{eq:disp0}. The wavy line is a brunch cut and the point $s=s_*$ is the reference point. We choose $s_*=2\pmass^2-(t/2)+i\mu$ ($\mu>0$).}
\label{fig:cont}
\end{figure}

In this study, we assume a weakly-coupled UV completion of gravity, whose illustrative example is perturbative string theory. The scattering amplitude will be then unitary and analytic order by order in perturbative expansions in terms of %the Planck scale 
$\Mpl$. Below, we simply write the $\phi\phi\to\phi\phi$ scattering amplitude up to $\mathcal{O}(\Mpl^{-2})$ as $\scat(s,t)$. Here, $(s,t,u)$ are Mandelstam variables satisfying $s+t+u=4\pmass^2$, where $\pmass^2$ is the pole mass of $\phi$.  
We assume that $\scat(s,t<0)$ is analytic in the complex $s$-plane except for discontinuities across the real $s$-axis, and it behaves mildly at high energies to satisfy $|\scat(s,t<0)/s^2|\to0$ in the limit $|s|\to\infty$ ($t$: fixed).~\footnote{In gapped systems, this mild behavior follows from the polynomial boundedness assumption~\cite {Froissart:1961ux, Martin:1962rt} in
  combination with the Phragm{\'{e}}n-Lindel{\"{o}}f theorem. Positivity bounds
  in the absence of the polynomial boundedness assumption are discussed in
  \cite {Tokuda:2019nqb}.} Then, the $s,u$-channel pole subtracted amplitude $\widetilde\scat(s,t)\coloneqq \scat(s,t)-(s,u\text{-poles})$ 
also satisfies the same properties. These assumptions lead to the relation (see also FIG.~\ref{fig:cont})
\begin{equation}
-\int_{\mathcal{C}_r}\frac{\mathrm{d}s'}{2\pi i}\frac{\widetilde\scat(s',t)}{\left(s'-s_*\right)^3}=\int_{\mathcal{C}_1+\mathcal{C}_2}\frac{\mathrm{d}s'}{2\pi i}\frac{\widetilde\scat(s',t)}{\left(s'-s_*\right)^3}\quad \text{for}\,t<0\,,\label{eq:disp0}
\end{equation}
where $s_*=2\pmass^2-(t/2)+i\mu$ and $\mu>0$. The contour $\mathcal{C}_r$ is a semi-circle centered at $s=s_*$ with a radius $r$. The contours $\mathcal{C}_1$ and $\mathcal{C}_2$ are straight lines defined by $\mathcal{C}_1\coloneqq\bigl\{s'|\,-\infty+i\mu<s'<s_*-r\bigr\}$ and $\mathcal{C}_2\coloneqq\bigl\{s'|\,s_*+r<s'<\infty+i\mu\bigr\}$, respectively. 
Next, we consider the low-energy expansion, $\widetilde\scat(s,t)=\sum_{n=0}^\infty \frac{c_{n}(t;\mu)}{n!}\left(s-s_*\right)^n+(t\text{-poles})$.
The residue of poles are polynomials and particularly we have $\der_s^2{\rm Res}_{t=0}\scat=-2\Mpl^{-2}$, ${\rm Res}_{t=0}\scat$ being the residue of the graviton $t$-channel pole. This reflects the spin-2 nature of graviton. The real part of eq.~\eqref{eq:disp0} in the limit $r\to+0$ then gives  
\begin{equation}
\re\,c_2(t;\mu)=\frac{32}{\pi}{\bf P}\int^\infty_{0}\mathrm{d}s'\frac{\im\,\widetilde{\scat}_{s}(s'+i\mu,t)}{\left(s'-u'\right)^3}+\frac{2}{\Mpl^2t}\label{eq:disp2}
\end{equation}
for $t<0$. Here, $u'\coloneqq 4\pmass^2-s'-t$ and ${\bf P}$ denoting the Cauchy principal value. We decomposed $\im\,\widetilde\scat$ into the $s$- and $u$-channel pieces as $\im\,\widetilde\scat(s,t)=\im\,\widetilde\scat_{s}(s,t)+\im\,\widetilde\scat_{u}(s,t)$ for $t<0$, and imposed the $s\leftrightarrow u$ crossing symmetry as $\im\,\widetilde\scat_{u}(u,t)=\im\,\widetilde\scat_{s}(s,t)$. 
We also used $\im\,\widetilde{\scat}_{s}(s,t)=0$ for $s<0$.  
One can evaluate the integral of $\im\,\widetilde\scat_{s}(s',t)$ at low energy regions $s'<\lth^2$ within EFT~\cite{Bellazzini:2016xrt, deRham:2017imi} to improve \eqref{eq:disp2} as\footnote{An importance of the improvement procedure done in \eqref{eq:impr1} to obtain non-trivial constraints on EFTs in the gravitational setups has been found recently in \cite{Alberte:2020jsk, Alberte:2020bdz}.}

\begin{align}
\impr(t)&\coloneqq\re\, c_2(t;\epsilon)-\frac{32}{\pi}{\bf P}\int^{\lth^2}_{0}\mathrm{d}s'\,\frac{\im\,\widetilde{\scat}_{s}(s'+i\epsilon,t)}{\left(s'-u'\right)^3}\no\\
&=\frac{32}{\pi}\int^\infty_{\lth^2}\mathrm{d}s'\,\frac{\im\,\widetilde{\scat}_{s}(s'+i\epsilon,t)}{\left(s'-u'\right)^3}+\frac{2}{\Mpl^2t}\,,
\label{eq:impr1}
\end{align}
where $\epsilon$ is an infinitesimal positive constant and $\lth$ denotes a threshold scale above which the EFT evaluation of $\scat$  
is unreliable: {\it e.g.}, we have $\lth\lesssim\Lambda$ when considering a model with a term $(\der\phi)^4/\Lambda^4$. We assume $\lth^2\gg \pmass^2$ throughout this study. The second term on the right-hand side (RHS) of \eqref{eq:impr1} diverges as $-\infty$ in the limit $t\to-0$. This makes the positivity of $\impr(0)$ unclear. In \cite{Tokuda:2020mlf}, the cancellation of $\mathcal{O}(t^{-1})$ terms on the RHS of \eqref{eq:impr1} is explicitly demonstrated by assuming the Regge behavior~\footnote{If we go beyond the $\mathcal{O}(\Mpl^{-2})$ analysis, the $\mathcal{O}(\ln^{-1}(s/\Ms^{2}))$ sub-leading correction to the Regge behavior plays an important role as discussed in~\cite{Herrero-Valea:2020wxz}. Also, we suppressed the positive contributions from other states which are irrelevant for the Reggeization of $t$-channel graviton exchange in \eqref{eq:regge2}.}
\begin{equation}
\im\,\widetilde\scat_{s}(s,t)\simeq f(t)\left(\frac{s}{\Ms^2}\right)^{2+j(t)}\left[1+\mathcal{O}\left(\frac{\Ms^2}{s}\right)\right]
+\cdots\,,\label{eq:regge2}
\end{equation}
at $s\gg\Ms^2$, where $f(t)$ and $j(t)$ are functions regular in the vicinity of $t=0$. A scale $\Ms$ denotes the lightest mass scale of the heavy physics which Reggeizes the amplitude. 
An explicit computation of the $\mathcal{O}(t^0)$ terms 
shows~\cite{Tokuda:2020mlf}
\begin{equation}
\impr(0)>-~\frac{\mathcal{O}(1)}{\Mpl^2\Ms^2}\,,\label{eq:scaling}
\end{equation}
assuming the single scaling $j'\sim |j''/j'|\sim |f'/f|\lesssim \mathcal{O}(\Ms^{-2})$ which is the case in tree-level amplitudes in perturbative string theory with $\Ms$ being the string scale. Here, the prime denotes the $t$-derivative evaluated at $t=0$.~\footnote{A similar order estimate of approximate positivity can be found in \cite{Hamada:2018dde, Loges:2020trf}. A proof of the single scaling is beyond the scope of this paper. The bound \eqref{eq:scaling} is distinct from the one conjectured in \cite{Alberte:2020jsk, Alberte:2020bdz} which depends on the EFT cutoff scales or mass scales of fields in EFTs.} Although the precise value of the RHS of \eqref{eq:scaling} will depend on the details of UV completion, this approximate positivity provides non-trivial constraints on EFTs as we shall see below.

\medskip
\section{III. Bounds on scalar theories coupled to gravity}
\subsection{A. Setup} Based on an inequality~\eqref{eq:scaling}, 
we shall derive a bound on a real scalar field theory coupled to gravity whose classical Lagrangian is given by eq.~\eqref{eq:action1},
with classical potential $V(\phi)$ and higher derivative terms $\lag_{\rm higher}$ of the form 
\begin{align*}
&V(\phi)=\frac{m^2\phi^2}{2}+\frac{g\phi^3}{3!}+\frac{\lambda\phi^4}{4!}+\cdots\,,\\
&\lag_{\rm higher}=\frac{\alpha(\der\phi)^4}{8\Lambda^4}+\cdots, 
\end{align*}
where $\alpha$ is a constant of order unity. Ellipses stand for higher order non-renormalizable terms which are present in general 
in the classical action because we regard this system as an EFT. Our choice of $\lag_{\rm higher}$ is necessary and sufficient for taking into account the influence of higher derivative corrections on positivity bounds up to $\mathcal{O}(\Lambda^{-4})$, thanks to the invariance of $\scat$ under perturbative field redefinitions.

We require $|g/m|, \,|\lambda|\lesssim1$ to ensure the perturbative expansion of $\scat$ in terms of coupling constants. We also require that the minimum of $V(\phi)$ is located at $\phi=0$ at least within the range $|\phi|\lesssim \lth$ to justify the perturbative evaluations of $\scat(s,t)$ up to $s\lesssim\lth^2$.

We use the dimensional regularization to regulate UV divergences and adopt the $\MSb$ scheme except we determine the counterterm of the form $Y\phi$ by imposing $\langle\phi\rangle=0$ (see appendices for details).

\begin{figure}
\includegraphics[width=100mm, trim=80 250 0 66]{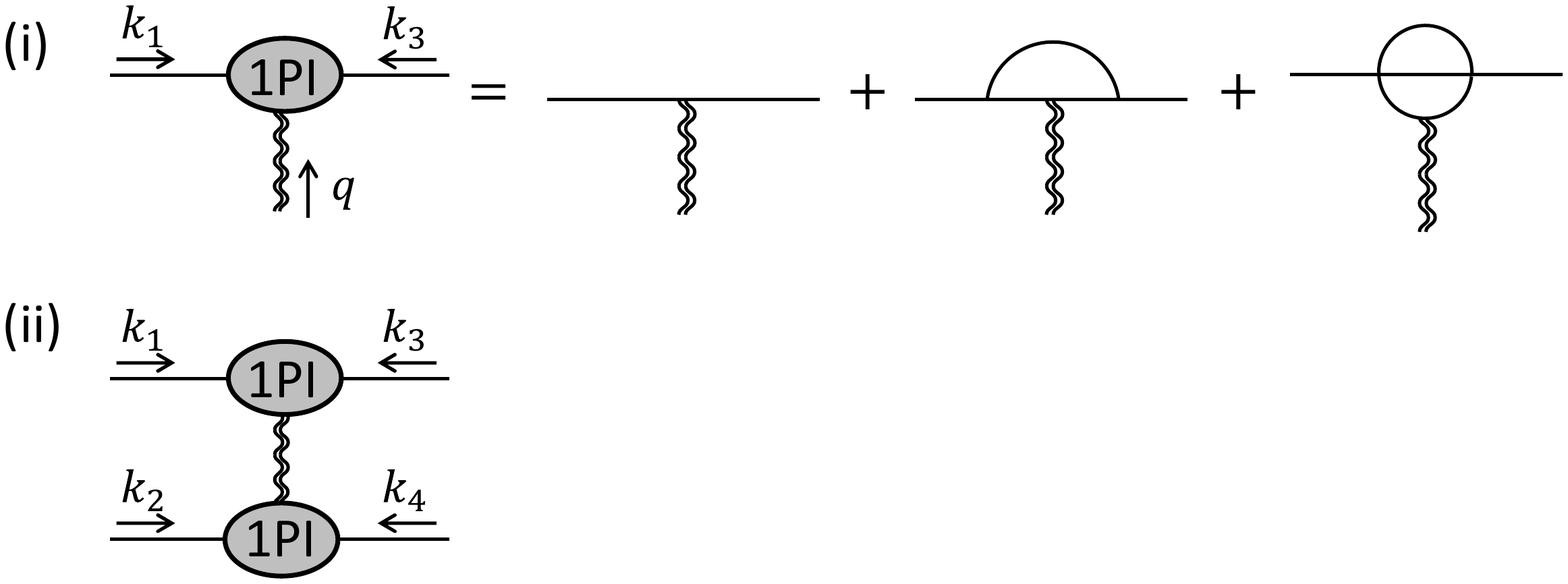}\caption{Diagrams (i): Relevant 1PI diagrams for an effective vertex $\phi^2 h$ in the present analysis. Solid lines and double wavy lines denote the propagators of $\phi$ and $h_{\mu\nu}$, respectively. Diagrams (ii): The $t$-channel diagrams which give negative contributions to $\impr(0)$, expressed in terms of the 1PI vertices shown in the diagrams (i).}
\label{fig:neg1}
\end{figure}

\medskip
\subsection{B. General bound} We compute $\impr(0)$. 
We may decompose $\impr(0)$ into the non-gravitational piece and the gravitational piece, $\impr(0)=\cngr+\cgr$.
Diagrams without gravitons and those with gravitons contribute to $\cngr$, and $\cgr$, respectively. 
We start with the non-gravitational one. The contributions from non-renormalizable terms such as $\alpha(\der\phi)^4/(8\Lambda^4)$ and $\phi^6/\Lambda^2$ can be written as $\tilde\alpha/\Lambda^4$. We have $\tilde\alpha=\alpha$ when considering a renormalizable potential. Renormalizable interactions also contribute to $\cngr$ via $s,u$-channel one-loop diagrams. Referring to the latter contributions as $\cngrp$, we have $\cngr=\frac{\tilde\alpha}{\Lambda^4}+\cngrp$ with 
\begin{eqnarray}
&&\cngrp\simeq\frac{\lambda^2}{16\pi^2\lth^4}+\frac{g^4}{12\pi^2m^2\lth^6}\no\\
&&\qquad\qquad\qquad-\frac{\lambda g^2}{6\pi^2\lth^6}\left(\ln\left(\frac{\lth^2}{m^2}\right)-\frac{1}{6}\right)\,.\label{eq:nongr1b}
\end{eqnarray}

When $\phi$ is a shift symmetric scalar, one can always make $\cngr$ positive by choosing $\alpha>0$, the coefficient of the higher derivative term $(\der\phi)^4$. This is reminiscent of the conventional positivity bound without gravity.

The presence of the gravitational piece $\cgr$ changes the story, however. In particular, loop corrections to the graviton $t$-channel exchange diagram give rise to negative contributions to which we refer as $\cgrt$: corresponding diagrams are shown in FIG.~\ref{fig:neg1}. Note that we are interested in amplitudes up to $\mathcal{O}(\Mpl^{-2})$, so that we can use the tree-level graviton propagator together with the loop-corrected one-particle irreducible (1PI) vertices.~\footnote{Note that negative contributions from the $t$-channel tree-level graviton exchange have been computed in different setups up to one-loop level \cite{Cheung:2014ega, Andriolo:2018lvp, Chen:2019qvr,Alberte:2020jsk, Alberte:2020bdz}.}
To compute $\cgrt$, we write the 1PI effective action $\Gamma$ as 
\begin{eqnarray}
&&\Gamma[\phi,h]\ni-\frac{1}{2}\int\frac{\mathrm{d}^4k}{(2\pi)^4}K(k^2)\phi(k)\phi(-k)\\
\no
&&+\frac{1}{4}\int\frac{\mathrm{d}^4k_1}{(2\pi)^4}\int\frac{\mathrm{d}^4k_3}{(2\pi)^4}V^{\mu\nu}\left(k_1,k_3\right)\phi(k_1)\phi(k_3)h_{\mu\nu}(q)\,,
\end{eqnarray}
where we suppressed terms irrelevant for our present analysis. Also, $g_{\mu\nu}=\eta_{\mu\nu}+h_{\mu\nu}$ and $q^\mu\coloneqq -(k_1+k_3)^\mu$. The kernel $K(k^2)$ denotes an inverse of a loop-corrected propagator of $\phi$, which is written in terms of the self-energy $\Pi(k^2)$ as $K(k^2)=k^2+m^2-\Pi(k^2)$. The 1PI $\phi\phi h$ vertex function is denoted by  $V^{\mu\nu}(k_1,k_3)$ and we parameterize it with imposing the on-shell conditions $k_1^2=k_3^2=-\pmass^2$ as 
\begin{eqnarray}
&&V^{\mu\nu}(k_1,k_3)|_{k_1^2=k_3^2=-\pmass^2}=T(q^2)\eta^{\mu\nu}+P(q^2)q^\mu q^\nu\no\\
&&\qquad\qquad\quad\!\!
-2Q(q^2)(p^\mu q^\nu+p^\nu q^\mu)+4R(q^2)p^\mu p^\nu\,.\label{eq:1PI}
\end{eqnarray}
Here, we defined $p^\mu\coloneqq(k_1-k_3)^\mu/2$. For example, at the tree-level approximation, we have $T_{\rm tree}=k_1.k_3-m^2$, $P_{\rm tree}=-1/2$, $Q_{\rm tree}=0$, and $R_{\rm tree}=1/2$. The final term on the RHS of \eqref{eq:1PI} expresses the transverse-traceless component of $h_{\mu\nu}$, and only this piece contributes to the spin-2 part of the $t$-channel graviton exchange diagrams which are represented by the lower diagram of FIG.~\ref{fig:neg1},
\begin{equation}
\scat(s,t)|_{{\rm FIG.~\ref{fig:neg1}}}=\frac{4R^2(-t)su}{\Mpl^2t}\times Z^2+\mathcal{O}(s^0)\,,
\end{equation}
where we used $q^2=-t$ and $Z$ is the residue of the propagator of $\phi$. We then find the relation $\cgrt=8Z^2\der_x(R^2(x))|_{x=0}/\Mpl^2\simeq 8R'(0)/\Mpl^2$ to get
\begin{equation}
\cgrt
\simeq-\frac{45-8\pi\sqrt{3}}{1296\pi^2}\frac{g^2}{\Mpl^2 m^4}-\frac{10-\pi^2}{4608\pi^4}\frac{\lambda^2}{\Mpl^2m^2}<0\,.\label{eq:grt}
\end{equation}
Note that the leading-order contributions from the $\phi^4$ vertex 
arise at the two-loop level.~\footnote{The appearance of the leading-order term at two-loop level is analogous to the fact that the leading-order correction to the field renormalization appears at two-loop level in $\lambda\phi^4$ theory.}
Also note that non-renormalizable terms in $V(\phi)$, such as $\phi^6/\Lambda^2$ vertex, contribute to $\cgrt$ at $\mathcal{O}(\Mpl^{-2}\Lambda^{-2})$, which are negligible compared to the $\mathcal{O}(\Lambda^{-4})$ contribution to $\cngr$ as long as $\Lambda^2\ll\Mpl^2$.
Interestingly, 
each term of \eqref{eq:grt} can be written in terms of the self-energy $\Pi$ as
\begin{equation}
\cgrt\simeq-\frac{2\left[\Pi''(-m^2)|_{\lambda=0}+2\Pi''(-m^2)|_{g=0}\right]}{3\Mpl^2}\,,\label{eq:grt2}
\end{equation}
where unitarity ensures $\Pi''(-m^2)>0$ because $\Pi(k^2)$ satisfies the twice-subtracted dispersion relation at least within the range of our approximation.
This implies that negativity of $\cgrt$ could be related to the physics of the loop-corrected self-energy. Although $\cgrt$ is determined by the behavior of the vertex $\phi^2h$ when the momentum of an external graviton is soft, it is not fixed by the soft graviton theorem alone: see appendices for details.

Diagrams other than those shown in FIG.~\ref{fig:neg1} also contribute to $\cgr$ at $\mathcal{O}(\Mpl^{-2})$. Such diagrams are the diagrams with $s,u$-channel tree-level graviton exchange diagrams, diagrams with a graviton-scalar conversion, and diagrams with a graviton propagator inside loops. We refer to these contributions as $\cgro$, which can be evaluated as
\begin{equation}
\cgro\sim\mathcal{O}\left(\frac{(g/m)^2}{\Mpl^2\lth^2},\frac{\lambda}{\Mpl^2\lth^2}\right)\,,\label{eq:est}
\end{equation}
and we have $\cgr=\cgrt+\cgro$. Practically, the term $\cgro$ can be ignored to read off the implication of the bound.~\footnote{This is because the $\mathcal{O}\bigl((g/m)^2/(\Mpl^2\lth^2)\bigr)$ terms are smaller than $|\cgrt|$ by factors of $m^2/\lth^2$, and the $\mathcal{O}\bigl(\lambda/(\Mpl^2\lth^2)\bigr)$ terms can be comparable to or larger than the $\mathcal{O}(\lambda^2/\lth^4)$ term contained in $\cngrp$ only when $\lambda/(\Mpl^2\lth^2)\lesssim\mathcal{O}(\Mpl^{-4})$.} We thus discuss the implications of an inequality,
\begin{equation}
\frac{\tilde\alpha}{\Lambda^4}+\cngrp+\cgrt>\frac{-\,\mathcal{O}(1)}{\Mpl^{2}\Ms^{2}}
\,.\label{eq:result2}
\end{equation}
Each term is given in eqs.~\eqref{eq:nongr1b} and \eqref{eq:grt}. Note that this bound is renormalization scheme independent at least within the range of our approximation. Also, we do not distinguish between $m$ and $\pmass$ because the difference comes in at higher orders.

\medskip
\subsection{C. Analysis of the bound}
\subsubsection{Emergence of a critical scale}
The bound \eqref{eq:result2} is meaningful only when the allowed negativity on the order of $\Mpl^{-2}\Ms^{-2}$ is negligible.  
We introduce {\it a critical energy scale} $\lcrit$ as $\lcrit\coloneqq(-\cgrt)^{-1/4}$, explicitly given as
\begin{equation}
\lcrit
=\left(\frac{10-\pi^2}{4608\pi^4}\frac{\lambda^2}{\Mpl^2m^2}+\frac{45-8\pi\sqrt{3}}{1296\pi^2}\frac{g^2}{\Mpl^2 m^4}\right)^{-\frac{1}{4}}\,,\label{eq:crit1}
\end{equation}
which is determined by the loop corrections to the self-energy in the present model as indicated by eq.~\eqref{eq:grt2}.
The $\mathcal{O}(\Mpl^{-2}\Ms^{-2})$ term can be ignored in eq.~\eqref{eq:result2} when the condition
\begin{equation}
{\rm Applicability \,\,Condition}:\,\, {\rm min}\left(\Lambda,\,\lcrit\right)<\sqrt{\Mpl\Ms}\label{eq:ac}
\end{equation}
is satisfied.
Under the condition \eqref{eq:ac}, we discuss the implications of gravitational positivity bounds \eqref{eq:result2}. It turns out that implications of \eqref{eq:result2} are clearly different between the following two cases,
\begin{align*}
{\rm Case\,(I):}\,\, \lcrit^4\gg\Lambda^4\,,\qquad {\rm Case\,(II):}\,\, \lcrit^4\ll\Lambda^4\,.
\end{align*}

\subsubsection{Case (I): Conventional positivity bounds} 
EFTs fall into this class when new physics appears well below the critical scale $\lcrit$. The bound~\eqref{eq:result2} on such models is well approximated by  
\begin{equation}
\frac{\tilde\alpha}{\Lambda^4}+\cngrp\geq0
\label{eq:bound1}\,.
\end{equation}
As explained below eq.~\eqref{eq:nongr1b}, this provides a constraint on non-renormalizable terms such as $\alpha(\der\phi)^4/(8\Lambda^4)$, and is the same as the conventional positivity bound without gravity. 
This is in accord with the decoupling of low-energy physics from the physics of quantum gravity.

\begin{figure}
\includegraphics[width=75mm, trim=160 50 10 150]{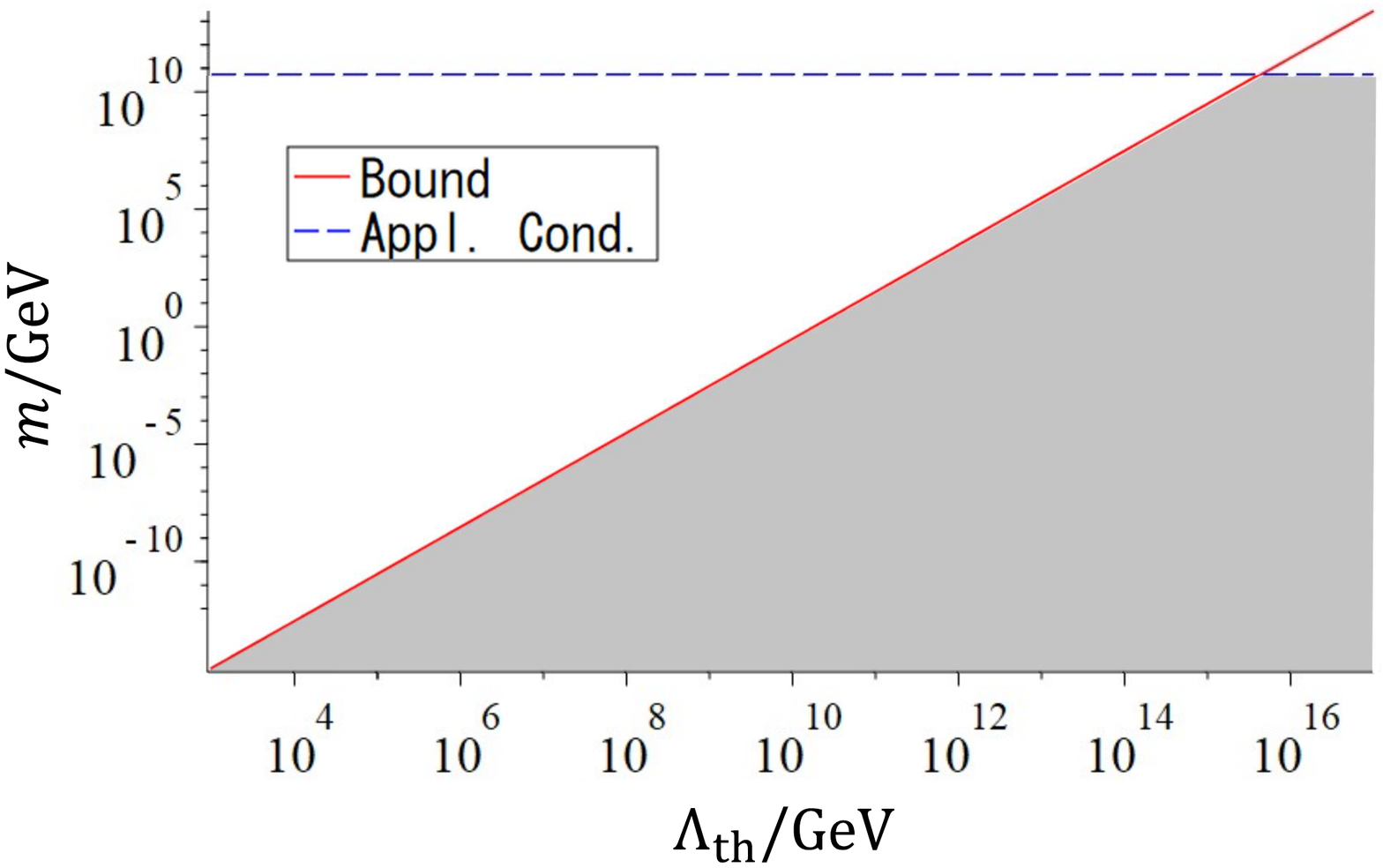}
\caption{Gravitational positivity bound on $\lambda\phi^4$ theory. Lines saturating our bound \eqref{eq:bound2} and the applicability condition \eqref{eq:ac} in $\lambda\phi^4$ theory are shown in the solid red line and dashes blue line, respectively. 
We substitute $\Ms=10^{16} {\rm GeV}$ and $\lambda=10^{-2}$ to draw the dashed lines. The shaded region is excluded by the bound \eqref{eq:bound2} under the condition \eqref{eq:lamac}.}
\label{fig:plot1}
\end{figure}
\subsubsection{Case (II): Bounds on scalar potentials} 
EFTs fall into this class when new physics appears well above the critical scale $\lcrit$.
The bound~\eqref{eq:result2} on such EFTs reads
\begin{equation}
\cngrp-\frac{1}{\lcrit^4}\geq0
\,.\label{eq:bound2}
\end{equation}
The left-hand side (LHS) of \eqref{eq:bound2} is determined once $V(\phi)$ is given, and hence eq.~\eqref{eq:bound2} constrains {\it the potential $V(\phi)$}. This is a genuinely new bound which is distinct from the ordinary positivity bound without gravity. 
The new bound \eqref{eq:bound2} typically prohibits an arbitrarily flat potential: for instance, in $\lambda\phi^4$ theory, eq.~\eqref{eq:bound2} gives {\it a lower bound} on $m^2$ for fixed $\lth$,
\begin{equation}
m\geq \sqrt{\frac{10-\pi^2}{288\pi^2}}\frac{\lth^2}{\Mpl}
\approx 2.8\times10^{9}\left(\frac{\lth}{10^{15}\,{\rm GeV}}\right)^2\,\,{\rm GeV}\,,\label{eq:lamb1}
\end{equation} 
while the applicability condition~\eqref{eq:ac} reads
\begin{equation}
m< 
5.4\times10^{12}\lambda\,\left(\frac{\Ms}{10^{16}\,{\rm GeV}}\right)\quad {\rm GeV}\,.\label{eq:lamac}
\end{equation}
Lines saturating eqs.~\eqref{eq:lamb1} and \eqref{eq:lamac} are plotted in the $(m,\,\lth)$-plane in FIG.~\ref{fig:plot1}. We find that the mass of the scalar field cannot be arbitrarily light in $\lambda\phi^4$ theory. In the presence of the cubic interaction, the expression of the bound \eqref{eq:bound2} is complicated, but it is obvious that tiny mass is prohibited for fixed $\lambda$ and $g/m$. In particular, when having the scaling $g^2\lesssim|\lambda|m^2$, the bound \eqref{eq:bound2} reads
\begin{equation}
m\gtrsim 
\frac{\lth^2}{\Mpl}\left[\frac{1.8\times10^{-2}~(g/m)^2}{\lambda^2}+ 4.6\times10^{-5}\right]^{1/2}
\,,\label{eq:mixbd}
\end{equation}
non-trivially constraining $V(\phi)$ for a given cutoff $\lth$. 

\subsection{D. Summary of the results}
To summarize, scalar potentials cannot be arbitrarily flat to be consistent with the gravitational positivity bound: for example, we cannot tune the mass to be arbitrarily smaller than the cutoff scale $\lth$ without violating the bound \eqref{eq:bound2}. Our result provides a quantitative swampland condition for scalar potentials which can be {\it derived} under several clear assumptions.

Any scalar field theory coupled to gravity which violates the bound \eqref{eq:bound2} has to possess appropriate non-renormalizable terms such as $\alpha(\der\phi)^4/(8\Lambda^4)$ with $\Lambda\lesssim\lcrit$ and $\alpha>0$, in order to satisfy the bound \eqref{eq:result2}. The presence of such non-renormalizable terms can be interesting phenomenologically. 
This is one of the main result of this study.

As a caveat, we remark however that our bound~\eqref{eq:result2} still has a room to accommodate models with a very tiny scalar mass. 
One can take the shift symmetric limit of a given massive scalar theory without violating the bound \eqref{eq:result2} by requiring $\lcrit\gg {\rm min}(\Lambda,\,\sqrt{\Mpl\Ms})$.~\footnote{Such a consistent shift symmetric limit can be explicitly realized in a model of a light scalar field whose tiny mass is protected by some symmetry. For instance, we consider the model of complex scalar $\Phi$ in which the approximate global $U(1)$ symmetry is spontaneously broken,
\begin{align*}
\lag=-\left|\der\Phi\right|^2-\frac{g}{4}\left(|\Phi|^2-\frac{v^2}{2}\right)^2+\frac{gv^2\epsilon}{2}\left(\Phi^2+{\Phi^*}^{2}\right)\,,
\end{align*}
with $g>0$, $v\neq0$, and $0<\epsilon\ll1$. One can check that the $\epsilon\to0$ limit provides a consistent shift symmetric limit for the pseudo Nambu-Goldstone boson.} For instance, a consistent shift symmetric limit of $\lambda\phi^4$ theory is $\lambda, m^2\to0$ with satisfying $|\lambda/m|\to0$.  It would be interesting if one could sharpen our analysis further to exclude all the flat potentials.

\medskip
\section{IV. Illustrative examples}
As an important application, we firstly discuss an implication of the bound \eqref{eq:bound2} to the renormalizable potential of the Higgs boson in the Standard Model. We then discuss more general form of potentials with non-renormalizable terms which have been frequently discussed in cosmology. We will consider axion-like particle models and the Starobinsky inflation as illustrative examples of non-renormalizable potentials.

\subsection{A. Renormalizable potential: Higgs boson}
Let us consider the implication of \eqref{eq:bound2} to the Higgs potential. The classical potential for the Higgs boson $\phi$ in the unitary gauge is parametrized as $m\sim 125$ GeV, $g/m\sim 1.5$, and $\lambda\sim 0.75$. For these values, the critical scale reads $\lcrit\sim10^{11}\,{\rm GeV}$. Then the applicability condition \eqref{eq:ac} reads $\Ms>10$ TeV, which is indeed satisfied in typical string theory scenarios.
So, it is reasonable to apply the gravitational positivity bound on the $2$ to $2$ scattering of the Higgs boson. Then, \eqref{eq:bound2} for the Higgs potential reads
\begin{align}
\lth\lesssim 1.9\sqrt{\Mpl m}\approx3.4\times 10^{10}\,\,\text{GeV}\,.\label{higgs}
\end{align}
Of course, it is necessary to include other Standard Model particles for more precise argument, but this result poses a non-trivial question if the Higgs sector of the Standard Model is in the Swampland. 
We will revisit this aspect in future work, which would open a new possibility to obtain non-trivial swampland constraints on the Standard Model coupled to gravity, particularly the Higgs sector.~\footnote{See also \cite{Aoki:2021ckh} for implications of gravitational positivity bounds on the light-by-light scattering in the Standard Model.}

\subsection{B. Non-renormalizable potentials}
Next, we consider the gravitational positivity bounds on potentials of the form
\begin{align}
V(\phi)=f^2m^2
\sum_{n=2}^\infty \frac{c_n}{n!} \left(\frac{\phi}{f}\right)^n\,, \label{genpot}
\end{align}
with $c_2=1$ and $|c_n|\lesssim \mathcal{O}(1)$ for $n=3,4,\cdots$. Here, $f$ denotes some energy scale satisfying $f\gg m$. Potentials of this form have been widely discussed in cosmology. For the potential \eqref{genpot}, the critical scale $\lcrit$ reads~\footnote{As we explained below eq.~\eqref{eq:grt}, non-renormalizable terms give negligible contributions to $\lcrit$.} 
\begin{equation}
\lcrit
=\left(\frac{10-\pi^2}{4608\pi^4}\frac{m^2}{f^2}c_4^2+\frac{45-8\pi\sqrt{3}}{1296\pi^2}c_3^2\right)^{-\frac{1}{4}}\sqrt{\Mpl f}\,,\label{crit2}
\end{equation}
implying that the bound \eqref{eq:bound2} is reliable only when $f\ll\Ms$ to satisfy the applicability condition $\lcrit<\sqrt{\Mpl\Ms}$. This potential is non-renormalizable unless $\{c_n\}_{\forall n\geq5}=0$. In the presence of non-renormalizable terms, we need to impose $\lth< f$. Under this assumption, we can ignore the contributions from non-renormalizable terms in $V(\phi)$. 

As an example, let us consider the case $|c_3|\sim|c_4|\sim \mathcal{O}(1)$ with $f\ll\Ms$. In this case, the bound \eqref{eq:bound2} reads 
\begin{align}
f\lesssim 7.4 \left|\frac{c_4}{c_3}\right|\left(\frac{m^2}{\lth^2}\right)\Mpl\ll\Mpl\,.\label{genbound2}
\end{align}
In the final line, we used $\lth^2\gg m^2$. This may be understood as a bound on the {\it flatness} of the potential because the potential becomes flatter for larger values of $f$. Note that for given $f$ and $m$, the bound \eqref{genbound2} can also be understood as an upper bound on the cutoff $\lth$.

\medskip
However, we do not always have non-trivial constraint on non-renormalizable potentials: models with non-renormalizable potentials \eqref{genpot} with $Z_2$ symmetry are always consistent with gravitational positivity bound. This is because, when the bound \eqref{eq:bound2} is violated in such models, we always find a super-Planckian critical scale $\lcrit$: 
\begin{align}
\lcrit&=\left(\frac{4608\pi^4}{10-\pi^2}\frac{f^2}{m^2c_4^2}\right)^{\frac{1}{4}}\sqrt{\Mpl f}\no\\
&>\sqrt{\frac{1152\pi^3}{c_4^2(10-\pi^2)}}\,\left(\frac{f}{\lth}\right)\Mpl>\Mpl\,.
\end{align}
 In the second line, we assume the violation of \eqref{eq:lamb1}. In the third line, we used $\lth<f$ and $|c_4|\lesssim\mathcal{O}(1)$. This analysis clarifies that the condition $\lth<f$ is a crucial obstruction to obtaining the bound on non-renormalizable potentials with $Z_2$ symmetry.
It would be interesting to embed such $Z_2$ symmetric potentials into renormalizable QFT models and study gravitational positivity bounds in these UV theories to derive non-trivial constraints on non-renormalizable potentials realized at low energies.
 
\subsubsection{Starobinsky inflation}
As a concrete example of phenomenologically relevant model with a non-renormalizable potential of the form \eqref{genpot}, we firstly consider the Starobinsky inflation model~\cite{Starobinsky:1980te} in which the potential of a scalar field minimally coupled to gravity is
\begin{equation}
V(\phi)=\frac{3}{4}\Mpl^2m^2\left[1-\exp\left(-\sqrt{\frac{2}{3}}\frac{\phi}{\Mpl}\right)\right]^2\,.
\end{equation} 
This potential takes the form of \eqref{genpot} with $f\sim\Mpl$. Then we find that the super-Planckian critical scale $\lcrit\approx 6.2\Mpl$, and hence we conclude that the Starobinsky inflation model is consistent with \eqref{eq:result2}.

\subsubsection{Axion-like particle}
Next, we consider a model of an axion-like particle whose potential is typically given by 
\begin{equation}
V(\phi)=f^2m^2[1-\cos(\phi/f)]\,,
\end{equation} 
where $f$ is the decay constant. This potential respects $Z_2$ symmetry and is non-renormalizable. We thus conclude that axion-like particle models  
are consistent with the gravitational positivity bound~\eqref{eq:result2}.

\medskip
These results suggest consistency between the well-motivated models and gravitational positivity bounds, which would support for the assumptions that we used to derive the gravitational positivity bounds.

\medskip
\section{V. Strong Scalar Weak Gravity Conjecture}
It is interesting to compare our bound \eqref{eq:bound2} with a bound called the Strong Scalar Weak Gravity Conjecture (SSWGC)~\cite{Gonzalo:2019gjp,Benakli:2020pkm,Gonzalo:2020kke}. 
For expansion coefficients of $V(\phi)$ around the vacuum at $\phi=0$, the SSWGC reads $|\xi(g/m)^2-\lambda|\geq(m^2/\Mpl^2)$ where $\xi$ is a constant of order unity.
In the absence of the quartic interaction, our bound \eqref{eq:bound2} reads 
\begin{equation}
\left(\frac{g}{m}\right)^2\geq r\,\frac{m^2}{\Mpl^2}\,,\quad r\approx0.014~\left(\frac{\lth^2}{m^2}\right)^3\gg1\,,
\end{equation}
which is stronger than the SSWGC bound because we have $\lth^2\gg m^2$ leading to $r\gg1$. Once turning on the quartic interaction, however, the implication of \eqref{eq:bound2} is different from but complementary to the one implied by the SSWGC. The SSWGC basically constrains $V(\phi)$ for tiny coupling constants $|\lambda|\sim(g/m)^2\sim\mathcal{O}(m^2/\Mpl^2)$: {\it e.g.}, the SSWGC reads $f<\Mpl$ for axion-like particles. By contrast, the bound \eqref{eq:bound2} non-trivially constrains $V(\phi)$ for larger coupling constants $|\lambda^2+(g/m)^2|\gtrsim (m^2/\Ms^2)$ because of the applicability condition. 

\medskip
 It would be interesting to study connections between the gravitational positivity and various conjectured bounds on $V(\phi)$.

\medskip
\section{VI. Conclusion} 
We derived a bound on scalar potentials by using the gravitational positivity bounds with clarifying assumptions and limitation of its applicability. 

\medskip
We identified the emergence of the critical energy scale $\lcrit$ which is determined in terms of coupling constants of renormalizable interactions. When the contributions of higher derivative terms can be ignored at the scale $\lcrit$, the gravitational positivity bound provides a genuinely new constraint \eqref{eq:bound2} on the potential $V(\phi)$. This is distinct from the ordinary positivity bounds in the absence of gravity. By contrast, 
the conventional positivity bounds for non-gravitational theories are recovered when some unknown heavy physics comes in well below the scale $\lcrit$ and the scales of quantum gravity $\Mpl$ and $\Ms$. This is in accord with the decoupling of low-energy physics from the physics of quantum gravity. 

\medskip
Interestingly, the critical scale $\lcrit$ can be much lower than the scales $\Mpl$ and $\Ms$. 
Any scalar theory coupled to gravity which violates the bound \eqref{eq:bound2} has to possess appropriate non-renormalizable terms such as $\alpha(\der\phi)^4/(8\Lambda^4)$ with $\Lambda\lesssim\lcrit$, $\alpha$ being a positive constant of order unity. Presence of such non-renormalizable terms can be phenomenologically interesting. It is particularly noteworthy that scalar potentials cannot be arbitrarily flat to be consistent with the bound \eqref{eq:bound2}: for instance, it is violated if we tune the mass to be much smaller than a given UV cutoff scale. 
This suggests the importance of the technical naturalness for embedding scalar theories into weakly coupled UV completion of gravity. 
Our result provides a quantitative swampland condition for scalar potentials which can be {\it derived} under several clear assumptions. 

\medskip
We also applied \eqref{eq:bound2} to the Higgs boson in the Standard Model and found a cutoff scale around $10^{10}$ GeV in \eqref{higgs}, which is much lower than the Planck scale. One cannot take this value seriously because we did not include the contributions from other Standard Model particles, but our result opens a new possibility to obtain non-trivial swampland constraints on the Standard Model coupled to gravity, particularly the Higgs sector. We leave this aspects for future work. 

\medskip
The essential origin of the presence of such non-trivial constraints is the negative sign of $\cgrt$ and the emergence of the critical scale $\lcrit$ at the scale much lower than the quantum gravity scale $\sqrt{\Mpl\Ms}$. We found that $\lcrit$ is determined by $\Pi''(-m^2)$ in the present analysis, $\Pi(k^2)$ being the self energy of $\phi$. This indicates that the emergence of $\lcrit$ and its value might be related to the physics of the loop-corrected self-energy. We leave further studies along this line of consideration for future work.

\medskip
\begin{acknowledgments}
\section{Acknowledgments}
We would like to thank Sota Sato for fruitful discussions and careful reading of the manuscript. We also would like to thank K. Benakli and E. Gonzalo for useful comments on the manuscript. T.N. is supported in part by JSPS KAKENHI Grant Numbers JP17H02894 and 20H01902, and MEXT KAKENHI Grant No. 21H00075.
J.\,T. is supported in part by JSPS Grants-in-Aid for Scientific Rersearch No. 202000912 and No.~21K13922.
\end{acknowledgments}

\medskip
\section{Appendix}

This appendix includes detailed computations which are omitted in the main text. In App.\,I, we compute $\Pi(k^2)$, the self-energy of $\phi$, which is used in (12). Computations of $\cngrp$, $\cgrt$, and $\cgro$ are shown in App.\,II, App.\,III, and App.\,IV, respectively. 

\section{Appendix I. Self-energy}
In this section, we compute the self-energy of $\phi$. To get UV-finite results, it is necessary to add counterterms. After adding the counterterms which are relevant for our analysis below, the action (1) becomes 
\begin{widetext}
\begin{align}
&S=\int\mathrm{d}^4x\,\sqrt{-g}\left[\frac{\Mpl^2}{2}R-\frac{1}{2}(\der\phi)^2-\frac{m^2}{2}\phi^2-\frac{g}{3!}\phi^3-\frac{\lambda}{4!}\phi^4-\Lag_{\rm ct}+\cdots\right]\,,\no\\
&\Lag_{\rm ct}=\frac{\delta Z_\phi}{2}(\der\phi)^2+Y\phi+\frac{\delta Z_m m^2}{2}\phi^2+\frac{\delta Z_gg}{3!}\phi^3+\frac{\delta Z_\lambda\lambda}{4!}\phi^4+\delta Z_{R\phi}R\phi+\delta Z_{R\phi^2}R\phi^2\,,\label{eq:action2}
\end{align}
\end{widetext}
where the ellipses stand for non-renormalizable terms which are irrelevant in the discussion below. Coefficients of counterterms are $\delta Z_\phi$, $Y$, $\delta Z_m$, $\delta Z_g$, $\delta Z_\lambda$, $\delta Z_{R\phi}$, and $\delta Z_{R\phi^2}$. To regulate UV divergences, we use the dimensional regularization and work in $d=4-\varepsilon$ dimensions. As the renormalization condition, we adopt the $\MSb$ scheme except we determine the counterterm $Y\phi$ by imposing $\langle\phi\rangle=0$. The value of $\impr(0)$ turns out to be scheme-independent at the level of approximation adopted in the present analysis.

\subsection{A. One-loop}
We begin by the one-loop analysis. Since gravitational corrections to the self-energy are sub-leading and irrelevant for our purpose in (12), our analysis here focuses on non-gravitational corrections\footnote{
 We will take care of gravitational corrections to the propagator appropriately, when we evaluate $\cgro$ in Sec. IV.}.
Firstly, the renormalization condition $\langle\phi\rangle=0$ leads to
\begin{align}
Y&=-\frac{g}{2}(-i)\tilde{\mu}^{4-d}\loopint i\Delta(\ell)\no\\
&=\frac{gm^2}{32\pi^2}\left(\frac{2}{\varepsilon}-\ln\left(\frac{m^2}{\mu^2}\right)+1\right)\,,\label{eq:cttadpole}
\end{align}
where $\Delta(k)$ denotes the free propagator of $\phi$ in momentum space: $
i\Delta(k)\coloneqq (k^2+m^2-i\epsilon)^{-1}$. Also, at the second equality, we defined $\mu^2\coloneqq 4\pi\tilde\mu^2\exp[-\gamma]$ with $\gamma$ being the Euler constant.
Next, we compute the self-energy. Relevant diagrams for the one-loop self-energy are shown in FIG.~\ref{fig:self_1}, which can be computed as 
\begin{widetext}
\begin{align}
\Pi_\text{one-loop}(k^2)&=-\frac{\lambda}{2}(-i)\,\tilde\mu^{4-d}\loopint i\Delta(\ell)+\frac{g^2}{2}(-i)\,\tilde\mu^{4-d}\loopint i\Delta(\ell)i\Delta(\ell+k)-\delta Z_mm^2-\delta Z_\phi k^2 \no\\
&=\frac{1}{16\pi^2\varepsilon}\left(m^2\lambda+g^2\right)-\frac{m^2\lambda}{32\pi^2}\left[\ln\left(\frac{m^2}{\mu^2}\right)-1\right]-\frac{g^2}{32\pi^2}\int^1_0\mathrm{d}x\,\ln\left(\frac{D_x(-k^2)}{\mu^2}\right)-\delta Z_mm^2-\delta Z_\phi k^2\,,
\end{align}
\begin{figure}[tbp]
  \centering
  \includegraphics[width=.5\textwidth, trim=150 400 300 135,clip]{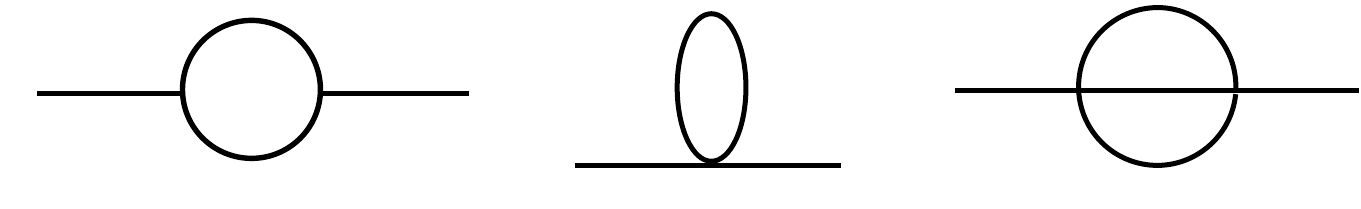}
 \caption{Diagrams relevant for the self-energy of $\phi$ in the present analysis. The first and the second diagrams are one-loop diagrams, and the third diagram is a two-loop diagram.
The second diagram is independent of the external momentum and so the leading corrections from the quartic coupling $\lambda$ to the momentum-dependence arise at the two-loop level.}
 \label{fig:self_1} 
\end{figure}
\end{widetext}
where $D_x(-k^2)\coloneqq (x^2-x)(-k^2)+m^2$. 
In the $\MSb$ scheme, we choose the counterterms as 
\begin{align}
\delta Z_m=\frac{\lambda+(g/m)^2}{16\pi^2\varepsilon}+\mathcal{O}(\lambda^2)\,,\quad \delta Z_\phi=\mathcal{O}(\lambda^2)\,,\label{eq:massct}
\end{align}
leading to
\begin{align}
\Pi_\text{one-loop}(k^2)=&-\frac{m^2\lambda}{32\pi^2}\left[\ln\left(\frac{m^2}{\mu^2}\right)-1\right]\no\\
&-\frac{g^2}{32\pi^2}\int^1_0\mathrm{d}x\,\ln\left(\frac{D_x(-k^2)}{\mu^2}\right)\,.
\end{align}
Then, at the one-loop level, the physical mass $\pmass$ is
\begin{align}
\pmass^2&=m^2-\Pi(-\pmass^2)\simeq m^2-\Pi_\text{one-loop}(-m^2)\no\\
&= m^2+\frac{m^2\lambda}{32\pi^2}\left[\ln\left(\frac{m^2}{\mu^2}\right)-1\right]\no\\
&\quad+\frac{g^2}{32\pi^2}\left[\ln\left(\frac{m^2}{\mu^2}\right)+\frac{\sqrt{3}\pi}{3}-2\right]\,,\label{eq:pmass}
\end{align}
and the inverse of the residue of the Feynman propagator of $\phi$ is 
\begin{align}
Z_\text{one-loop}^{-1}&=1-\Pi'_\text{one-loop}(-\pmass^2)\no\\
&\simeq 1-\Pi'_\text{one-loop}(-m^2)\no\\
&=1+\frac{\left(2\pi\sqrt{3}-9\right)g^2}{288\pi^2m^2}\,.\label{eq:residue}
\end{align}
For later convenience, we list up the value of $\Pi'$, $\Pi''$, and $\Pi'''$:
\begin{subequations}
\label{oneloopresult}
\begin{align}
&\Pi'_\text{one-loop}(-m^2)=\frac{g^2}{288\pi^2m^2}\left(9-2\pi\sqrt{3}\right)\,,\label{eq:one1}\\
&\Pi''_\text{one-loop}(-m^2)=\frac{g^2}{864\pi^2m^4}\left(45-8\pi\sqrt{3}\right)>0\,,\label{eq:one2}\\
&\Pi'''_\text{one-loop}(-m^2)=\frac{g^2}{216\pi^2m^6}\left(27-5\pi\sqrt{3}\right)\,.\label{eq:one3}
\end{align}
\end{subequations}
Using $\im\,\Pi_\text{one-loop}(k^2-i\epsilon)|_{k^2\leq-4m^2}=\frac{g^2}{32\pi}\sqrt{\frac{-4m^2-k^2}{-k^2}}$, we find the twice-subtracted dispersion relation~\cite{Srednicki:2007qs}
\begin{align}
\Pi''_\text{one-loop}(-m^2)=\frac{2}{\pi}\int^\infty_{4m^2}\mathrm{d}s\,\frac{\im\,\Pi_\text{one-loop}(-s-i\epsilon)}{(s-m^2)^3}>0\,,\label{twice}
\end{align}
implying that the positivity of $\Pi''_\text{one-loop}(-m^2)$ is ensured by unitarity.

\subsection{B. Two-loop}
Next, we compute the $\mathcal{O}(\lambda^2)$ correction to $\Pi(k^2)$, which is at the two-loop level.
This is practically important because the $\mathcal{O}(\lambda)$ corrections to $\Pi(k^2)$ are $k$-independent and so $k$-dependent corrections from the quartic coupling $\lambda$ first arise from two-loop diagrams: see the third diagram shown in FIG.~\ref{fig:self_1}. The contribution from this diagram reads 
\begin{align}
&\Pi_\text{two-loop}(k^2)|_{g=0}\no\\
&=\frac{\lambda^2}{6}\left((-i)\tilde\mu^{4-d}\right)^2\loopint\int\frac{\mathrm{d}^dq}{(2\pi)^d}\no\\
&\qquad\quad\times i\Delta(\ell)i\Delta(q)i\Delta(q+\ell-k)\no\\
&\quad+\text{($k$-independent diagrams)}\,,
\end{align}
where the second term is for $k$-independent diagrams (double-scoop diagrams) that are not relevant for the following discussion. To evaluate the double-integral in the first term, it is convenient to insert $1=\frac{1}{2d}(\frac{\partial\ell_\mu}{\partial\ell_\mu}+\frac{\partial q_\mu}{\partial q_\mu})$ as
\begin{align}
&\loopint\int\frac{\mathrm{d}^dq}{(2\pi)^d}\,i\Delta(\ell)i\Delta(q)i\Delta(q+\ell-k)\no\\
&=
\loopint\int\frac{\mathrm{d}^dq}{(2\pi)^d}\,
\frac{1}{2d}\left(\frac{\partial \ell_\mu}{\partial \ell_\mu}+\frac{\partial q_\mu}{\partial q_\mu}\right)\no\\
&\qquad\times
i\Delta(\ell)i\Delta(q)i\Delta(q+\ell-k)
\,.
\end{align}
Reformulating the right hand side by partial integrals, we find
\begin{subequations}
\label{eq:2loop1}
\begin{align}
&\Pi_\text{two-loop}(k^2)|_{g=0}=\frac{\lambda^2}{6(3-d)}\left(3m^2K(k^2)+k_\mu K^\mu(k^2)\right)\no\\
&\qquad\qquad\qquad\qquad\,\,+\text{($k$-independent diagrams)}\,,\\
&K(k^2)\coloneqq\left((-i)\tilde\mu^{4-d}\right)^2\loopint\int\frac{\mathrm{d}^dq}{(2\pi)^d}\no\\
&\qquad\qquad\qquad\times i\Delta(\ell)\left[i\Delta(q)\right]^2i\Delta(q+\ell-k)\,,\\
&K^\mu(k^2)\coloneqq\left((-i)\tilde\mu^{4-d}\right)^2\loopint\int\frac{\mathrm{d}^dq}{(2\pi)^d}\no\\
&\qquad\qquad\qquad\times i\Delta(\ell)\left[i\Delta(q)\right]^2i\Delta(q+\ell-k)q^\mu\,.
\end{align}
\end{subequations}
To compute $K$ and $K^\mu$, we firstly perform the integration over $\ell$ by using the Feynman integral formula. We then 
perform the integration over $q$ by using the formula again to get
\begin{subequations}
\label{eq:2loop2}
\begin{align}
&K(k^2)=\frac{\Gamma(4-d)}{(4\pi)^d\left(\frac{d}{2}-2\right)}\int^1_0\mathrm{d}x\,(x-x^2)^{(d/2)-2}\no\\
&\qquad\quad\times\int^1_0\mathrm{d}y\, y^{2-(d/2)}\,\frac{\mathrm{d}}{\mathrm{d}y}\left[(1-y)\left(\frac{\tilde\mu^2}{F_{xy}(-k^2)}\right)^{4-d}\right]\,,\\
&K^\mu(k^2)= k^\mu \frac{\Gamma(4-d)}{(4\pi)^d}\int^1_0\mathrm{d}x\,(x-x^2)^{(d/2)-2}\no\\
&\qquad\qquad\times\int^1_0\mathrm{d}y\, y^{2-(d/2)}(1-y)\left(\frac{\tilde\mu^2}{F_{xy}(-k^2)}\right)^{4-d}\,,
\end{align}
\end{subequations}
where $F_{xy}(-k^2)\coloneqq(y-y^2)k^2+\left[(1-y)+\frac{y}{x-x^2}\right]m^2$.
To arrive at the above expressions \eqref{eq:2loop2}, we also used $d<4$. Notice that double integrals in these expressions are regular even in the $d\to4$ limit\footnote{We learned this trick in the QFT lecture by Atsuo Kuniba held at  the University of Tokyo -- Komaba, when one of the authors was a PhD student. We thank him for giving nice lectures.}. 
\begin{widetext}
Substituting eqs.~\eqref{eq:2loop2} into \eqref{eq:2loop1} and expanding the resultant expressions in terms of an infinitesimal positive parameter $\varepsilon$, we have
\begin{align}
\Pi_{\rm two\text{-}loop}(k^2)|_{g=0}&= \frac{1}{3}\left(\frac{\lambda}{16\pi^2}\right)^2\times\Biggl\{ -\frac{3m^2}{\varepsilon^2}+\frac{1}{\varepsilon}\left[3m^2\left(\ln\left(\frac{m^2}{\mu^2}\right)-\frac{3}{2}\right)-\frac{k^2}{4}\right]\no\\
&\!\quad
-3m^2\left[\frac{9}{4}-2\ln\left(\frac{m^2}{\mu^2}\right)+\frac{1}{2}\ln^2\left(\frac{m^2}{\mu^2}\right)-\frac{1}{2}\int^1_0\mathrm{d}x\int^1_0\mathrm{d}y\,\ln y\,\frac{\mathrm d}{\mathrm dy}\left((1-y)\ln\,\biggl(\frac{\mu^2}{F_{xy}(-k^2)}\biggr)\right)+\frac{\pi^2}{24}\right]\no\\
&\!\quad
-k^2\left[\frac{5}{16}+\frac{1}{2}\int^1_0\mathrm{d}x\int^1_0\mathrm{d}y\,(1-y)\ln\,\biggl(\frac{\mu^2}{F_{xy}(-k^2)}\biggr)\right]\Biggr\}
+\text{($k$-independent diagrams)}\,.\label{eq:2loop3}
\end{align}
\end{widetext}
UV-divergent terms are shown in the first line. The terms in the second and the third lines are UV finite. For our purpose, it is enough to compute $\Pi'$ and $\Pi''$. We renormalize the UV divergent terms proportional to $k^2$ in \eqref{eq:2loop3} by choosing the field renormalization $\delta Z_\phi$ appropriately to obtain the UV finite expression for $\Pi'$, while $\Pi''$ is UV finite and independent of the renormalization scheme. The $\MSb$ choice of $\delta Z_\phi$ is 
\begin{align}
\delta Z_\phi=-\frac{1}{12\varepsilon}\left(\frac{\lambda}{16\pi^2}\right)^2\,,\label{eq:fdct}
\end{align}
leading to 
\begin{subequations}
\begin{align}
&\Pi'_\text{two-loop}(-m^2)|_{g=0}=\frac{1}{12}\left(\frac{\lambda}{16\pi^2}\right)^2\left[
\ln\left(\frac{m^2}{\mu^2}\right)+\frac{3}{4}\right]\,,\label{eq:two1}\\
&\Pi''_\text{two-loop}(-m^2)|_{g=0}=\frac{1}{24m^2}\left(\frac{\lambda}{16\pi^2}\right)^2\left(10-\pi^2\right)>0\,.\label{eq:two2}
\end{align}
\end{subequations}

\section{Appendix II. Computation of $\cngrp$}
In this section, we compute $\cngrp$ by evaluating non-gravitational scattering amplitudes generated by renormalizable self-interactions of $\phi$.
The leading contributions are through the one-loop
diagrams shown in FIG.~\ref{fig:nongrav_diag1} and FIG.~\ref{fig:self_2}. We define their sum by $\scat_{\rm non\text{-}grav,\,ren}=\scat_{(a)}+\scat_{(b)}+\scat_{(c)}+\scat_{(d)}+\scat_{(e)}$.
These diagrams consist of renormalizable vertices only, and hence $\scat_{\rm non\text{-}grav,\,ren}$ is analytic in the complex $s$-plane modulo poles and cuts, and satisfies the Froissart bound. 
Then, the following relation holds: 
\begin{align}
&\re \,c_2(0;\epsilon)|_{\rm non\text{-}grav,\,ren}\no\\
&=\frac{4}{\pi}\int^\infty_{4\pmass^2}\mathrm{d}s'\,\frac{\im\,\scat_{\rm non\text{-}grav,\,ren}(s',0)}{(s'-2\pmass^2)^3}\,.\label{eq:ngdis}
\end{align}
We can explicitly check this equality by directly computing both sides of \eqref{eq:ngdis}, although we do not show detailed computations here. Eq.~\eqref{eq:ngdis} leads to  
\begin{align}
\cngrp=\frac{4}{\pi}\int^\infty_{\lth^2}\mathrm{d}s'\,\frac{\im\,\scat_{\rm non\text{-}grav,\,ren}(s',0)}{(s'-2\pmass^2)^3}\,.\label{eq:ngren1}
\end{align}
The imaginary part of each diagram can be computed as 
\begin{widetext}
\begin{subequations}
\label{eq:ngrim1}
\begin{align}
&\im\,\scat_{\rm(a)}(s,0)|_{s\geq 4m^2}
=\frac{\lambda^2}{32\pi}\sqrt{\frac{s-4m^2}{s}}\,,
\qquad\im\,\scat_{(b)}(s,0)|_{s\geq 4m^2}
=\frac{-\lambda g^2}{8\pi}\sqrt{\frac{1}{s(s-4m^2)}}\,\ln\left(\frac{s-3m^2}{m^2}\right)\,,\\
&\im\,\scat_{(c)}(s,0)|_{s\geq 4m^2}
=\frac{g^4}{16\pi m^2}\sqrt{\frac{s-4m^2}{s}}\frac{1}{(s-3m^2)}+\frac{g^4}{8\pi}\frac{1}{\sqrt{s(s-4m^2)}(s-2m^2)}\ln\left(\frac{s-3m^2}{m^2}\right)\,,\\&
\im\,\scat_{(d)}(s,0)|_{s\geq 4m^2}=\frac{g^4}{32\pi}\frac{1}{(m^2-s)^2}\sqrt{\frac{s-4m^2}{s}}\,,\\
&\im\,\scat_{\rm (e)}(s,0)|_{s\geq 4m^2}=\frac{\lambda g^2}{16\pi}\sqrt{\frac{s-4m^2}{s}}\frac{1}{s-m^2}+\frac{g^4}{8\pi}\frac{1}{\sqrt{s(s-4m^2)}(m^2-s)}\ln\left(\frac{s-3m^2}{m^2}\right)\,.
\end{align}
\end{subequations}
\begin{figure}[tbp]
 \centering
  \includegraphics[width=.7\textwidth, trim=120 370 100 80,clip]{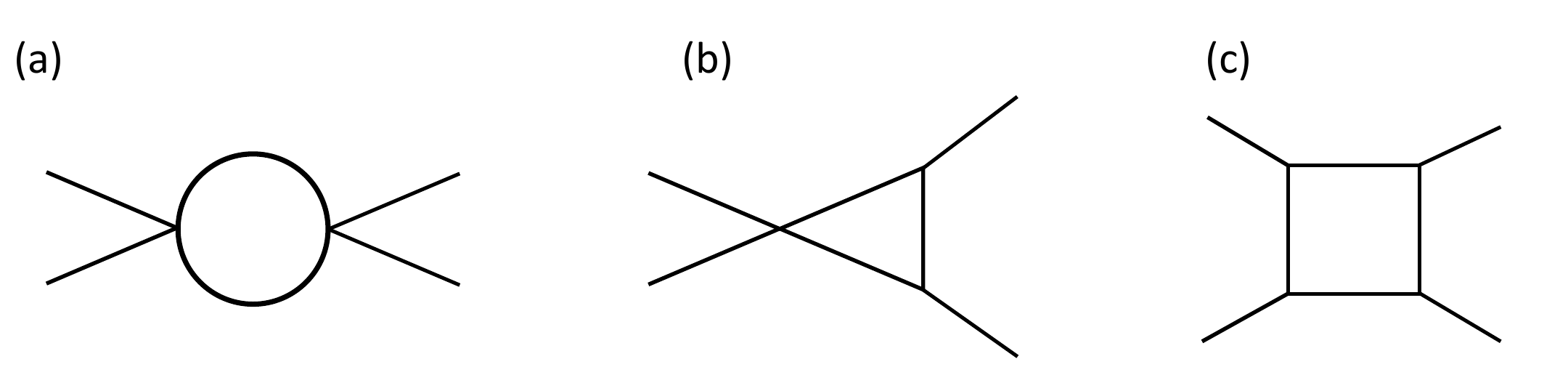}
 \caption{1PI diagrams for non-gravitational four-point scattering up to $\mathcal{O}(\lambda^2,\lambda g^2, g^4)$.  All the possible assignments of external momenta should be considered.}
 \label{fig:nongrav_diag1} 
\end{figure}
\begin{figure}[tbp]
 \centering
  \includegraphics[width=.8\textwidth, trim=120 285 100 95,clip]{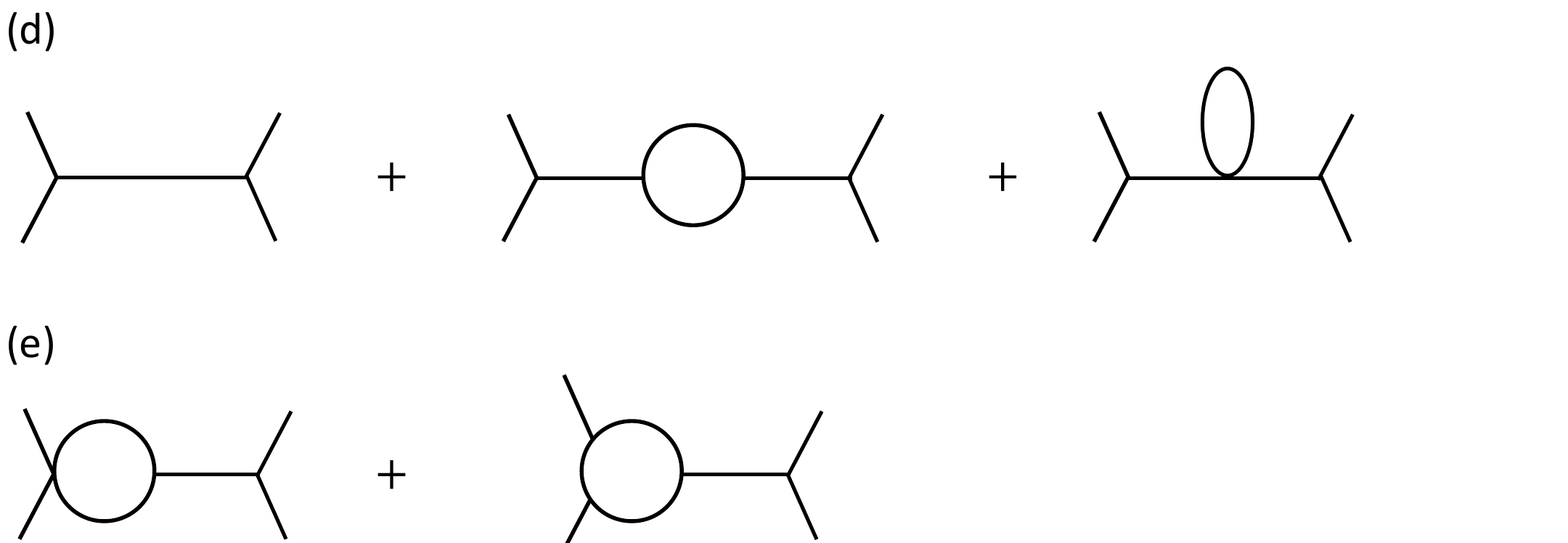}
 \caption{The process $\phi\phi\to\phi\phi$ expressed by the non-1PI diagrams up to one-loop.  All the possible assignments of external momenta should be considered. Diagrams (d) represent the self-energy corrections to the $\phi\phi\to\phi\phi$ process. The diagram (e) is the contribution from the one-loop corrections to $\phi^3$ vertex.}
 \label{fig:self_2} 
\end{figure}
Here, we set $m^2=\pmass^2$ since the difference comes in at higher orders in the coupling constants. We do not encounter any UV divergences in these computations, implying the scheme-independence of $\cngrp$ at the leading order. Substituting eqs.~\eqref{eq:ngrim1} into \eqref{eq:ngren1}, we obtain 
\begin{align}
\cngrp\simeq\frac{\lambda^2}{16\pi^2\lth^4}+\frac{g^4}{12\pi^2m^2\lth^6}-\frac{\lambda g^2}{6\pi^2\lth^6}\left(\ln\left(\frac{\lth^2}{m^2}\right)-\frac{1}{6}\right)\,,\label{eq:nogrposi1}
\end{align}
where tiny corrections suppressed by $m^2/\lth^2$ are discarded.

\section{Appendix III. Computation of $\cgrt$}
In this section we compute $\cgrt$ by evaluating the $t$-channel graviton exchange diagram with loop-corrected vertices. Since the computation is somewhat involved, it is helpful to use the Ward-Takahashi (WT) identities associated with the general covariance for consistency check and also for simplifying the calculation. Note that we distinguish between $\pmass^2$ and $m^2$ in Sec.~A, although we set $m^2=\pmass^2$ in Sec.~B  since the difference comes in at higher orders in the coupling constants.
\subsection{A. Ward-Takahashi identities}
Firstly, we derive the WT identities associated with the general covariance. Suppose that the 1PI effective action of $\phi$ and the metric fluctuation $h_{\mu\nu}$ defined by $g_{\mu\nu}=\eta_{\mu\nu}+h_{\mu\nu}$ is of the form,
\begin{align}
\Gamma&=-\frac{1}{2}\int \frac{d^4k}{(2\pi)^4}K(k^2) \phi(-k)\phi(k)+\frac{1}{4}\int \frac{d^4k_1}{(2\pi)^4}\int \frac{d^4k_2}{(2\pi)^4}
V^{\mu\nu}(k_1,k_2)\phi(k_1)\phi(k_2)h_{\mu\nu}(-k_1-k_2)+\cdots\,,
\end{align}
\end{widetext}
where the dots stand for terms that are irrelevant for graviton exchange diagrams studied in the main text and also for the WT identities derived in this section. For example, for the tree-level action, we have
\begin{align}
\label{tree_gravity}
&K_{\rm tree}(k^2)=m^2+k^2\,,\no\\
&V^{\mu\nu}_{\rm tree}(k_1,k_2)=-\eta^{\mu\nu}(m^2-k_1.k_2)-(k_1^\mu k_2^\nu+k_2^\mu k_1^\nu)\,.
\end{align}
The effective action should be invariant under the infinitesimal transformation,
\begin{align}
&\delta \phi(k)=i\int\frac{d^4k_2}{(2\pi)^4}\epsilon_\mu(k-k_2)k_2^\mu\phi(k_2)\,,\no\\
&\delta h_{\mu\nu}(k)=ik_\mu\epsilon_\nu(k)+ik_\nu\epsilon_{\mu}(k)+\mathcal{O}(h)\,,
\end{align}
so that we require
\begin{align}
\label{WT_pre}
0&=i\int \frac{d^4k_1}{(2\pi)^4}\int \frac{d^4k_2}{(2\pi)^4}
\phi(k_1)\phi(k_2)\epsilon_\mu(-k_1-k_2)\no\\
&\times\left[
-K(k_1^2)k_2^\mu-K(k_2^2)k_1^\mu
-
(k_1+k_2)_\nu V^{\mu\nu}(k_1,k_2)
\right]\,.
\end{align}
This is of course satisfied for the tree-level action~\eqref{tree_gravity}. Now
let us assume that $V^{\mu\nu}$ is local and so it can be expanded in the momenta $k_i$. Under this assumption, we may write $V^{\mu\nu}$ as
\begin{align}
&V^{\mu\nu}(k_1,k_2)=T(k_1,k_2)\eta^{\mu\nu}+P(k_1,k_2)q^\mu q^\nu\no\\
&\quad-2Q(k_1,k_2)\left(p^\mu q^\nu+p^\nu q^\mu\right)+4R(k_1,k_2)p^\mu p^\nu\,,
\end{align}
where $T$, $P$, $Q$, and $R$ are local scalar functions, and we introduced $p^\mu$ and $q^\mu$ by $p^\mu\coloneqq(k_1-k_2)^\mu/2$ and $q^\mu\coloneqq-(k_1+k_2)^\mu$. For example, for the tree-level action, we have
\begin{align}
&T_{\rm tree}(k_1,k_2)=k_1.k_2-m^2\,,\quad P_{\rm tree}(k_1,k_2)=-\frac{1}{2}\,,\no\\
&Q_{\rm tree}(k_1,k_2)=0\,,\quad R_{\rm tree}(k_1,k_2)=\frac{1}{2}\,.
\end{align}
Also note that when the graviton $h_{\mu\nu}$ is on-shell, i.e., when $h_{\mu\nu}$ is transverse traceless, only the $R$ component contributes to the amplitude. In this language, the constraint~\eqref{WT_pre} reads
\begin{widetext}
\begin{align}
0&=-q^\mu\left[
T(k_1,-k_1-q)
+q^2P(k_1,-k_1-q)-\left(q^2+2(k_1.q)\right)Q(k_1,-k_1-q)
+\frac{K(k_1^2)+K\left(k_1^2+2(k_1.q)+q^2\right)}{2}
\right]
\nonumber
\\
&\quad
+2p^\mu \left[q^2Q(k_1,-k_1-q)-\left(q^2+2(k_1.q)\right)R(k_1,-k_1-q)+\frac{-K(k_1^2)+K\left(k_1^2+2(k_1.q)+q^2\right)}{2}\right]\,,
\end{align}
which leads to the following two identities:
\begin{subequations}
\begin{align}
0&=T(k_1,-k_1-q)
+q^2P(k_1,-k_1-q)-\left(q^2+2(k_1.q)\right)Q(k_1,-k_1-q)
+\frac{K(k_1^2)+K\left(k_1^2+2(k_1.q)+q^2\right)}{2}\,,\label{constraint1}
\\
0&=q^2Q(k_1,-k_1-q)-\left(q^2+2(k_1.q)\right)R(k_1,-k_1-q)+\frac{-K(k_1^2)+K\left(k_1^2+2(k_1.q)+q^2\right)}{2}\,.\label{constraint2}
\end{align}
\end{subequations}
They are the WT identities associated with the general covariance.
\end{widetext}

\medskip
We then discuss the consequence of these identities in the soft limit $q\to0$ while $k_1$ being fixed.
For this purpose, we introduce the soft limit expansion,
\begin{align}
A(k_1,-k_1-q)=\sum_{n,\,m=0}^\infty\Delta_{n,\,m} A(k_1^2)\,(k_1.q)^n(q^2)^m\,,\label{eq:exp1}
\end{align}
where $A=T,P,Q,R$. In terms of these expansion coefficients, the WT identity \eqref{constraint2} reads 
\begin{subequations}
\label{eq:soft}
\begin{align}
\mathcal{O}(q)\,\,: \quad&\Delta_{0,0}R(k_1^2)=\frac{1}{2}\left(1-\Pi'(k_1^2)\right)\,,\label{eq:wt4}\\
\mathcal{O}(q^2): \quad&\Delta_{0,0}Q(k_1^2)=0\,,\quad\Delta_{1,0}R(k_1^2)=\frac{-1}{2}\Pi''(k_1^2)\,,\label{eq:wt5}\\
\mathcal{O}(q^3):\quad&\Delta_{1,0}Q(k_1^2)-2\Delta_{0,1}R(k_1^2)=\frac{1}{2}\Pi''(k_1^2)\,,\no\\
&\Delta_{2,0}R(k_1^2)=\frac{-1}{3}\Pi'''(k_1^2)\label{eq:wt6}\,,
\end{align}
\end{subequations}
up to $\mathcal{O}(q^3)$. Here, we used the fact that $K(k^2)$ can be written in terms of the self-energy $\Pi(k^2)$ as $
K(k^2)=k^2+m^2-\Pi(k^2)$. 
We can also derive identities that involve $T$ from the constraint~\eqref{constraint1}: the result is 
\begin{subequations}
\label{eq:soft2}
\begin{align}
\mathcal{O}(q^0)\,\,: \quad&\Delta_{0,0}T(k_1^2)=-\left(k_1^2+m^2-\Pi(k_1^2)\right)\,,\label{eq:wt7}\\
\mathcal{O}(q)\,\,: \quad&\Delta_{1,0}T(k_1^2)=-\frac{1}{2}\left(1-\Pi'(k_1^2)\right)\,,\label{eq:wt8}\\
\mathcal{O}(q^2): \quad&\Delta_{2,0}T(k_1^2)-2\Delta_{1,0}Q(k_1^2)=\Pi''(k_1^2)\,,\no\\
&\Delta_{0,1}T(k_1^2)+\Delta_{0,0}P(k_1^2)=-\frac{1}{2}\left(1-\Pi'(k_1^2)\right)\,,\label{eq:wt9}\\
\mathcal{O}(q^3):\quad&\Delta_{3,0}T(k_1^2)-2\Delta_{0,2}Q(k_1^2)=-\frac{2}{3}\Pi'''(k_1^2)\,,\no\\
&\Delta_{1,1}T(k_1^2)+\Delta_{1,0}P(k_1^2)\no\\
&-\left(\Delta_{1,0}Q(k_1^2)+2\Delta_{0,1}Q(k_1^2)\right)=\Pi''(k_1^2)\,.\label{eq:wt10}
\end{align}
\end{subequations}
Here, we used the first identity of \eqref{eq:wt5} to derive eq.~\eqref{eq:wt8} and the second equality of \eqref{eq:wt9}. 

\medskip
Before moving on to concrete loop computations, we summarize implications of the WT identities for $R'(0)$ that is relevant for the evaluation of $\cgrt$ (recall discussion around equations~(9)-(11)). The function $R(t)$ defined in~(9) is given in the present language as
\begin{align}
&R(q^2)= R(k_1,-k_1-q)|_{k_1^2=k_2^2=-\pmass^2}
\nonumber
\\
&=\sum_{n,\,m=0}^\infty\Delta_{n,m} R(-\pmass^2)\left(\frac{-q^2}{2}\right)^n\left(q^2\right)^m\no\\
&=\Delta_{0,0}R(-\pmass^2)+\Bigl[\Delta_{0,1}R(-\pmass^2)-\frac{1}{2}\Delta_{1,0}R(-\pmass^2)\Bigr]q^2\no\\
&\quad+\mathcal{O}\left(q^4\right)
\,,
\end{align}
where we used the relation $k_1.q=-q^2/2$ that holds when $k_1^2=k_2^2$. Correspondingly, we have 
\begin{align}
R'(0)&=\Delta_{0,1}R(-\pmass^2)-\frac{1}{2}\Delta_{1,0}R(-\pmass^2)\no\\
&=\frac{1}{2}\Delta_{1,0}Q(-\pmass^2)\,,\label{eq:wtcoeff2}
\end{align}
where we used~\eqref{eq:soft} at the second equality. Note that the WT identities relate $R'(0)$ directly to $\Delta Q_{1,0}(-\pmass^2)/2$, but its sign cannot be fixed from the symmetry consideration alone. It would be interesting to provide a physical interpretation of the sign of $\Delta Q_{1,0}(-\pmass^2)/2$, leaving it for future work.

\subsection{B. WT identities at one-loop}

\begin{figure}[tbp]
 \centering
  \includegraphics[width=.47\textwidth, trim=40 225 400 80,clip]{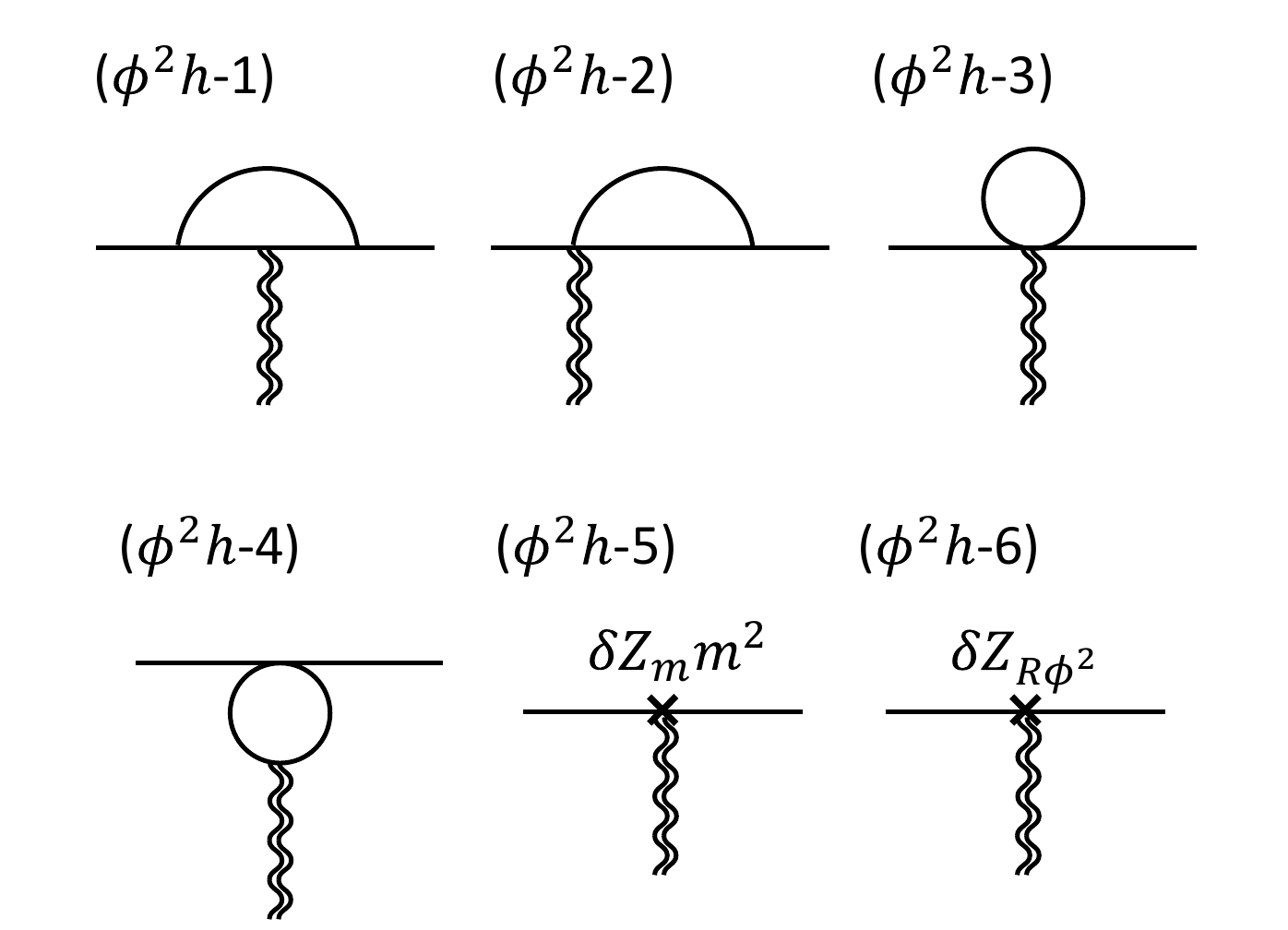}
 \caption{One-loop 1PI diagrams for the $\phi^2 h$-vertex. Counterterm diagrams are also shown which are necessary for renormalization. The diagram with a field renormalization $\delta Z_\phi$ is not included since it is not necessary at one-loop level in the present model. External in-going momenta for scalar lines are $k_1$ and $k_2$. In-going momentum for external graviton is $q=-(k_1+k_2)$. All the possible assignment of external momenta should be considered.}
 \label{fig:1PI_hpp} 
\end{figure}

Now let us perform loop computations. For consistency check, we begin by computing off-shell $Q$ and $R$ at the one-loop level  and demonstrating that they indeed satisfy the WT identities~\eqref{eq:soft}.
The one-loop 1PI diagrams relevant for the $\phi\phi h$-vertex are shown in FIG.~\ref{fig:1PI_hpp}. In this figure, in-going momenta for external scalar lines are referred to as $k_1$ and $k_2$. The in-going momentum for the external graviton is $q=-(k_1+k_2)$. All diagrams in FIG.~\ref{fig:1PI_hpp} contribute to the trace part $T(k_1,k_2)$, whereas one-loop corrections to $Q(k_1,-k_1-q)$ and $R(k_1,-k_1-q)$ arise only from the diagram $(\phi^2 h\text{-}1)$.
One-loop corrections to the vertex function from the diagram $(\phi^2 h\text{-}1)$ can be written as 
\begin{align}
V^{\mu\nu}_{(\phi^2 h\text{-}1)}(k_1,k_2)=g^2&\tilde\mu^{4-d}(-i)\loopint V^{\mu\nu}_{\rm tree}(-\ell,\ell-q)\no\\
&\times i\Delta(\ell)i\Delta(\ell-q)i\Delta(\ell+k_1)\,.\label{eq:1loop1}
\end{align}
Then, the one-loop corrections to the off-shell vertex function $Q$ and $R$ are 
\begin{align}
&Q_\text{one-loop}\left(k_1,-k_1-q\right)\no\\
&\quad=\frac{g^2}{32\pi^2}\int^1_0\mathrm{d}x\,\int^{1-x}_0\mathrm{d}y\,\frac{-x(1-x-2y)}{D_{xy}(k_1,q)}\,,\label{eq:off1}\\
&R_\text{one-loop}\left(k_1,-k_1-q\right)\no\\
&\quad=\frac{g^2}{32\pi^2}\int^1_0\mathrm{d}x\,\int^{1-x}_0\mathrm{d}y\,\frac{x^2}{D_{xy}(k_1,q)}\,,\label{eq:off2}
\end{align}
where
\begin{align}
D_{xy}(k_1,q)&\coloneqq (y-y^2)q^2+2xy\left(k_1.q\right)+(x-x^2)k_1^2+m^2\,.\label{eq:defd1}
\end{align}
Note that $Q$ and $R$ are UV finite and so we set $d=4$ in the above. On the other hand, the terms $T$ and $P$ have UV divergent terms. The diagrams $(\phi^2 h\text{-}1)$, $(\phi^2 h\text{-}2)$, and $(\phi^2 h\text{-}3)$ give UV divergent terms to  $\Delta_{0,0}T$ which can be renormalized by the mass renormalization $\delta Z_m$ given by \eqref{eq:massct}. The diagram $(\phi^2 h\text{-}4)$ gives UV divergences to $\Delta_{0,1}T$ and $\Delta_{0,0}P$ with an opposite sign which can be renormalized by $\delta Z_{R\phi^2}$. 

\medskip
Using explicit one-loop results \eqref{oneloopresult}, \eqref{eq:off1}, and \eqref{eq:off2}, we demonstrate that eqs.~\eqref{eq:soft} are indeed satisfied. We basically show the results when $k_1^2=-m^2$ below for simplicity, but one can easily extend the computation to $k_1^2\neq -m^2$ case at least when $0<-k_1^2<4m^2$. We start with computing $\Delta_{0,0}R_\text{one-loop}(-m^2)$,
\begin{widetext}
\begin{align}
\Delta_{0,0}R_\text{one-loop}(-m^2)&=\frac{g^2}{32\pi^2 m^2}\int^1_0\mathrm{d}x\,\int^{1-x}_0\mathrm{d}y\,\frac{x^2}{(x^2-x+1)}=\frac{-g^2}{32\pi^2 m^2}\left(\frac{1}{2}-\frac{\pi\sqrt{3}}{9}\right)=\frac{-1}{2}\Pi'_\text{one-loop}(-m^2)\,,
\end{align}
confirming eq.~\eqref{eq:wt4}.  Next, we confirm \eqref{eq:wt5} by
\begin{align}
&\Delta_{0,0}Q_\text{one-loop}(-m^2)=\frac{-g^2}{32\pi^2 m^{2}}\int^1_0\mathrm{d}x\,\int^{1-x}_0\mathrm{d}y\,\frac{x(1-x-2y)}{x^2-x+1}=0\,,\\
&\Delta_{1,0}R_\text{one-loop}(-m^2)=\frac{-g^2}{32\pi^2 m^{4}}\int^1_0\mathrm{d}x\,\int^{1-x}_0\mathrm{d}y\,\frac{2x^3y}{\left(x^2-x+1\right)^{2}}=\frac{-g^2}{32\pi^2 m^{4}}\left[\frac{5}{6}-\frac{4\sqrt{3}\pi}{27}\right]=\frac{-1}{2}\Pi''_\text{one-loop}(-m^2)\,.
\end{align}
Also, $\Delta_{1,0}Q_\text{one-loop}(-m^2)$ and $\Delta_{0,1}R_\text{one-loop}(-m^2)$ are computed as 
\begin{align}
\Delta_{1,0}Q_\text{one-loop}(-m^2)&=\frac{-g^2}{32\pi^2 m^4}\int^1_0\mathrm{d}x\,\int^{1-x}_0\mathrm{d}y\,\frac{-2x^2y(1-x-2y)}{(x^2-x+1)^2}=\frac{-1}{6}\Pi''_\text{one-loop}(-m^2)\,,\label{eq:diq1}\\
\Delta_{0,1}R_\text{one-loop}(-m^2)&=\frac{g^2}{32\pi^2 m^4}\int^1_0\mathrm{d}x\,\int^{1-x}_0\mathrm{d}y\,\frac{x^2(y^2-y)}{(x^2-x+1)^2}=\frac{-1}{3}\Pi''_\text{one-loop}(-m^2)\,,
\end{align}
leading to the first identity of \eqref{eq:wt6}:
\begin{align}
\Delta_{1,0}Q_\text{one-loop}(-m^2)-2\Delta_{0,1}R_\text{one-loop}(-m^2)=\frac{1}{2}\Pi''_\text{one-loop}(-m^2)\,.
\end{align}
Finally, we confirm the second identity of \eqref{eq:wt6} by computing $\Delta_{2,0}R_\text{one-loop}(-m^2)$:
\begin{align}
\Delta_{2,0}R_\text{one-loop}(-m^2)&=\frac{g^2}{32\pi^2 m^6}\int^1_0\mathrm{d}x\,\int^{1-x}_0\mathrm{d}y\,\frac{4x^4y^2}{(x^2-x+1)^3}=\frac{-g^2}{8\pi^2 m^6}\left(\frac{1}{3}-\frac{5\pi\sqrt{3}}{81}\right)=\frac{-1}{3}\Pi'''_\text{one-loop}(-m^2)\,.
\end{align}

\subsection{C. Computation of $R'(0)$}
We compute $R'(0)$, which is relevant for the positivity bounds as shown in eq.~(11), by evaluating $\Delta_{1,0}Q(-m^2)$ and using \eqref{eq:wtcoeff2}. To efficiently compute $\Delta_{1,0}Q(-m^2)$, we perform the soft expansion of the integrand before performing the loop integrals. We firstly check how our method works at the one-loop level. We then compute two-loop corrections to  $\Delta_{1,0}Q(-m^2)$ from the quartic coupling $\lambda$, leading to the second term of (11). Final results are eqs.~\eqref{eq:r1} and \eqref{eq:r2}, confirming that $\cgrt$ is given by eqs.~(11) and (12). It also shows the renormalization scheme independence of $\cgrt$.
\subsubsection{1. One-loop corrections}
One-loop corrections to the vertex function from the diagram $(\phi^2 h\text{-}1)$ is computed as \eqref{eq:1loop1}. Since we are interested in $\Delta_{1,0}Q(-m^2)$, we can discard the terms proportional to $\eta^{\mu\nu}$ in $V^{\mu\nu}_{\rm tree}(-\ell,\ell-q)$. That is, what we have to compute is
\begin{align}
V^{\mu\nu}_{(\phi^2 h\text{-}1)}(k_1,k_2)\ni g^2\tilde\mu^{4-d}(-i)\loopint \left[\ell^\mu(\ell-q)^\nu+\ell^\nu(\ell-q)^\mu\right]i\Delta(\ell)i\Delta(\ell-q)i\Delta(\ell+k_1)\,.\label{eq:softQ1}
\end{align}
Next, to simplify the computation, we use the fact that it is enough to consider the soft limit $q\to 0$ with fixing $k_1$ and imposing the on-shell condition on the external graviton momentum, $q^2=0$. In this limit, we do not need to distinguish between $k_1.q$ and $p.q$, because $k_1.q=p.q$ when $q^2=0$.
We consider the soft-expansion of $i\Delta(\ell-q)$ with imposing $q^2=0$, 
\begin{align}
i\Delta(\ell-q)=i\Delta(\ell)\left[1+2i\Delta(\ell)(\ell.q)+\left[2i\Delta(\ell)(\ell.q)\right]^2+\mathcal{O}((\ell.q)^3)\right]\,.
\end{align}
Using this expansion, we obtain the $\mathcal{O}((\ell.q),(\ell.q)^2)$ terms of the RHS of eq.~\eqref{eq:softQ1} as 
\begin{align}
\left.{\rm RHS\,\,of}\,\,\eqref{eq:softQ1}\right|_{\mathcal{O}((\ell.q),(\ell.q)^2)~{\rm terms}}&=6g^2\int^1_0\mathrm{d}x\,(1-x)^2\tilde\mu^{4-d}(-i)\loopint \frac{\left[\ell^\mu(\ell-q)^\nu+\ell^\nu(\ell-q)^\mu\right](\ell.q)}{\left[(\ell+xk_1)^2+D_x(-k_1^2)\right]^{4}}\no\\
&\quad
\times\left\{1+\frac{8(1-x)(\ell.q)}{3\left[(\ell+xk_1)^2+D_x(-k_1^2)\right]}\right\}+\mathcal{O}((\ell.q)^3)\,.
\end{align}
In terms of $L\coloneqq \ell+xk_1=\ell+x(p-(q/2))$, we have
\begin{align}
(\ell.q)|_{\ell=L-xp+\frac{xq}{2}}&=(\ell.q)-x(p.q)\,,\\
\ell^\mu(\ell-q)^\nu+\ell^\nu(\ell-q)^\mu|_{\ell=L-xp+\frac{xq}{2}}&=2L^\mu L^\nu-2x\left(L^\mu p^\nu+L^\nu p^\mu \right)+(x-1)\left(L^\mu q^\nu+L^\nu q^\mu \right)\no\\
&\quad
+(x-x^2)\left(p^\mu q^\nu+p^\nu q^\mu\right)+2x^2p^\mu p^\nu+\left(\frac{x^2}{2}-x\right)q^\mu q^\nu\,.
\end{align} 
Here, we used $q^2=0$. By using the fact that we can perform the following replacement in the integrand,
\begin{align}
&\left(L^\mu p^\nu+L^\nu p^\mu \right)(\ell.q)\rightarrow \frac{L^2}{d}\left(p^\mu q^\nu+ p^\nu q^\mu \right)\,,
\end{align}
we can compute the one-loop corrections to $\Delta_{1,0}Q(k_1^2)$ as 
\begin{align}
\Delta_{1,0} Q_\text{one-loop}(k_1^2)&=-3g^2\int^1_0\mathrm{d}x \,x^2(1-x)^3\tilde\mu^{4-d}(-i)\int\frac{\mathrm{d}^dL}{(2\pi)^d}\left[\frac{-1}{\left(L^2+D_x(-k_1^2)\right)^4}+\frac{8L^2}{3\left(L^2+D_x(-k_1^2)\right)^5}\right]\no\\
&=
-\frac{g^2}{96\pi^2}\int^1_0\mathrm{d}x\,\frac{x^2(1-x)^3}{\left[D_x(-k_1^2)\right]^2}\,.
\end{align}
In particular, we impose $k_1^2=-m^2$ to get
\begin{align}
\Delta_{1,0} Q_\text{one-loop}(-m^2)= \frac{-g^2}{32\pi^2 m^4}\left(\frac{5}{18}-\frac{4\pi\sqrt{3}}{81}\right)=\frac{-1}{6}\Pi''_\text{one-loop}(-m^2)<0\,.
\end{align}
This precisely coincides with the result \eqref{eq:diq1} which is obtained from \eqref{eq:off1}, the full off-shell expression of one-loop corrections to $Q(k_1,-k_1-q)$ before taking the soft limit.  We conclude by using eq.~\eqref{eq:wtcoeff2} that the leading order contribution to $R'(0)$ arises at one-loop level in the presence of the cubic coupling $g$, 
\begin{align}
R'_\text{one-loop}(0)=\frac{-1}{12}\Pi''_\text{one-loop}(-m^2)=\frac{-g^2}{64\pi^2 m^4}\left(\frac{5}{18}-\frac{4\pi\sqrt{3}}{81}\right)<0\,.\label{eq:r1}
\end{align}
It is found that $R'_\text{one-loop}(0)$ as well as $\Pi''_\text{one-loop}$ is independent of the renormalization scheme.

\subsubsection{2. Two-loop corrections}

We have seen that the quartic coupling $\lambda$ does not affect $R'(0)$ at the one-loop level and its leading contribution appears at the two-loop level.
We thus compute the two-loop corrections to $Q(k_1,k_2)$ with setting $g=0$ to obtain $R'(0)$ via the relation~\eqref{eq:wtcoeff2}. Diagrams relevant to $R'(0)$ are shown in FIG.~\ref{fig:1PI_hpp_2}.
We refer to their contribution as $V^{\mu\nu}_{{\rm FIG.}~\ref{fig:1PI_hpp_2}}$, which is given by
\begin{align}
V^{\mu\nu}_{{\rm FIG.}~\ref{fig:1PI_hpp_2}}&=\frac{\lambda^2}{2}\left((-i)\tilde\mu^{4-d}\right)^2\loopint\loopintt\,i\Delta(\ell)i\Delta(\ell-q)i\Delta(r)i\Delta(\ell+r+k_1)V^{\mu\nu}_{\rm tree}(-\ell,\ell-q)\no\\
&\quad
-\delta Z_\phi\left[\left(k_1^\mu k_2^\nu+k_1^\nu k_2^\mu\right) -\eta^{\mu\nu}(k_1.k_2)\right]\,.\label{eq:tloopq0}
\end{align}
To compute $\Delta_{1,0}Q(-m^2)$, we can also discard the terms  proportional to $\eta^{\mu\nu}$ in $V^{\mu\nu}_{\rm tree}(-\ell,\ell-q)$. It is then enough for us to compute 
\begin{align}
V^{\mu\nu}_{{\rm FIG.}~\ref{fig:1PI_hpp_2}}&\ni\frac{\lambda^2}{2}\left((-i)\tilde\mu^{4-d}\right)^2\loopint\loopintt\,i\Delta(\ell)i\Delta(\ell-q)i\Delta(r)i\Delta(\ell+r+k_1)\left[\ell^\mu(\ell-q)^\nu+\ell^\nu(\ell-q)^\mu\right]\no\\
&\quad
-\delta Z_\phi\left(k_1^\mu k_2^\nu+k_1^\nu k_2^\mu\right) \,.\label{eq:tloopq1}
\end{align}
Next, we expand $i\Delta(\ell-q)$ in terms of the soft momentum $q$ of graviton with imposing the on-shell condition $q^2=0$, and we perform the integration over $r$. The result up to $\mathcal{O}((\ell.q)^2)$ is
\begin{align}
{\rm RHS\,\,of}\,\,\eqref{eq:tloopq1}
&=\frac{\lambda^2}{2}\frac{\Gamma(2-\frac{d}{2})}{(4\pi)^{\frac{d}{2}}}(-i)\left(\tilde\mu^{4-d}\right)^2\loopint\left[i\Delta(\ell)\right]^2 \left[1+2(\ell.q)i\Delta(\ell)+4(\ell.q)^2\left[i\Delta(\ell)\right]^2\right]\no\\
&\quad
\times\left[\ell^\mu(\ell-q)^\nu+\ell^\nu(\ell-q)^\mu\right]\int^1_0\mathrm{d}x\frac{1}{\left[D_x(-(\ell+k_1)^2)\right]^{2-\frac{d}{2}}}-\delta Z_\phi\left(k_1^\mu k_2^\nu+k_1^\nu k_2^\mu\right)+\mathcal{O}((\ell.q)^3)\,.\label{eq:tloopq2}
\end{align}
\begin{figure}[tbp]
 \centering
  \includegraphics[width=.35\textwidth, trim=120 370 400 80,clip]{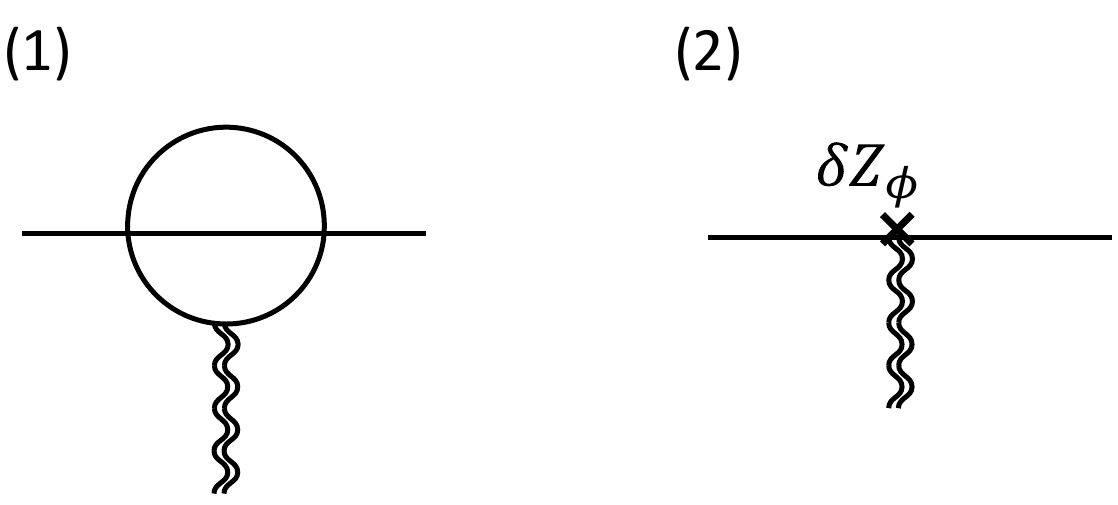}
 \caption{Two-loop 1PI diagrams for the $\phi^2 h$-vertex from the quartic coupling $\lambda$ which are relevant for the computation of $Q(k_1,k_2)$ and $R(k_1,k_2)$. Other diagrams are relevant only for the computation of $T(k_1,k_2)$, the trace component of $V^{\mu\nu}$. External in-going momenta for scalar lines are $k_1$ and $k_2$ and that for the graviton is referred to as $q=-(k_1+k_2)$.}
 \label{fig:1PI_hpp_2} 
\end{figure} 
To perform the integration over $\ell$, it is useful to note
\begin{align}
&(1-y)(\ell^2+m^2)+yD_x(-(\ell+k_1)^2)
=\left[1-y(x^2-x+1)\right]\left[L'^2+E_{xy}(-k_1^2)\right]\,,\\
&L'\coloneqq\ell+\alpha k_1\,,\quad \alpha\coloneqq \frac{yx(1-x)}{1-y(x^2-x+1)}\,,\quad E_{xy}(-k_1^2)\coloneqq (\alpha-\alpha^2)k_1^2+\frac{m^2}{1-y(x^2-x+1)}\,.
\end{align}
Then, by using the Feynman's integral formula as usual, 
the RHS of eq.~\eqref{eq:tloopq2} can be computed as 
\begin{align}
&{\rm RHS\,\,of}\,\,\eqref{eq:tloopq1}\no\\
&=\frac{\lambda^2}{2}\frac{\Gamma(4-\frac{d}{2})}{(4\pi)^{\frac{d}{2}}}(-i)\left(\tilde\mu^{4-d}\right)^2\int^1_0\mathrm{d}x\int^1_0\mathrm{d}y\,\frac{(1-y)y^{1-\frac{d}{2}}}{\left[1-y(x^2-x+1)\right]^{4-\frac{d}{2}}}\loopint\frac{\left[\ell^\mu(\ell-q)^\nu+\ell^\nu(\ell-q)^\mu\right]}{\left[L'^2+E_{xy}(-k_1^2)\right]^{4-\frac{d}{2}}}\no\\
&\quad
+\frac{\lambda^2}{2}\frac{\Gamma(5-\frac{d}{2})}{2(4\pi)^{\frac{d}{2}}}(-i)\left(\tilde\mu^{4-d}\right)^2\int^1_0\mathrm{d}x\int^1_0\mathrm{d}y\,\frac{(1-y)^{2}y^{1-\frac{d}{2}}}{\left[1-y(x^2-x+1)\right]^{5-\frac{d}{2}}}\loopint\frac{2(\ell.q)\left[\ell^\mu(\ell-q)^\nu+\ell^\nu(\ell-q)^\mu\right]}{\left[L'^2+E_{xy}(-k_1^2)\right]^{5-\frac{d}{2}}}\no\\
&\quad
+\frac{\lambda^2}{2}\frac{\Gamma(6-\frac{d}{2})}{6(4\pi)^{\frac{d}{2}}}(-i)\left(\tilde\mu^{4-d}\right)^2\int^1_0\mathrm{d}x\int^1_0\mathrm{d}y\,\frac{(1-y)^{3}y^{1-\frac{d}{2}}}{\left[1-y(x^2-x+1)\right]^{6-\frac{d}{2}}}\loopint\frac{4(\ell.q)^2\left[\ell^\mu(\ell-q)^\nu+\ell^\nu(\ell-q)^\mu\right]}{\left[L'^2+E_{xy}(-k_1^2)\right]^{6-\frac{d}{2}}}\no\\
&\quad
-\delta Z_\phi\left(k_1^\mu k_2^\nu+k_1^\nu k_2^\mu\right)+\mathcal{O}(q^3)\,.\label{eq:tloopq3}
\end{align}
Terms in the first three lines are contributions from the diagram (1) in FIG.~\ref{fig:1PI_hpp_2}: the first, second, and third lines are of order $(\ell.q)^0$, $(\ell.q)^1$, and $(\ell.q)^2$, respectively. The counterterm diagram (2) is responsible for the term in the final line. To identify the terms which contribute to $\Delta_{1,0}Q(k_1^2)$, we shall use 
\begin{align}
(\ell.q)|_{\ell=L'-\alpha p+\frac{\alpha q}{2}}&=(L'.q)-\alpha(p.q)\,,\label{eq:inner1}\\
\ell^\mu(\ell-q)^\nu+\ell^\nu(\ell-q)^\mu|_{\ell=L'-\alpha p+\frac{\alpha q}{2}}&=2L'^\mu L'^\nu-2\alpha\left(L'^\mu p^\nu+L'^\nu p^\mu \right)+(\alpha-1)\left(L'^\mu q^\nu+L'^\nu q^\mu \right)\no\\
&\quad+(\alpha-\alpha^2)\left(p^\mu q^\nu+p^\nu q^\mu\right)+2\alpha^2p^\mu p^\nu+\left(\frac{\alpha^2}{2}-\alpha\right)q^\mu q^\nu\,,\label{eq:inner2}
\end{align} 
and perform the following replacement in the integrand of the RHS of \eqref{eq:tloopq3}:
\begin{align}
&\left(L'^\mu p^\nu+L'^\nu p^\mu \right)(L'.q)\rightarrow \frac{L'^2}{d}\left(p^\mu q^\nu+ p^\nu q^\mu \right)\,.
\end{align}
To get the first two relations \eqref{eq:inner1} and \eqref{eq:inner2}, we used $q^2=0$. Only the $\mathcal{O}((\ell.q),(\ell.q)^2)$ terms of the RHS of \eqref{eq:tloopq3} are relevant for the computation of $\Delta_{1,0}Q_\text{two-loop}|_{g=0}$:
the result is
\begin{align}
\Delta_{1,0}Q_\text{two-loop}(k_1^2)|_{g=0}
&=\frac{-1}{12}\left(\frac{\lambda}{16\pi^2}\right)^2\int^1_0\mathrm{d}x\int^1_0\mathrm{d}y\,\frac{y(1-y)^3x^2(1-x)^2}{\left[1-y(x^2-x+1)\right]^6E_{xy}(-k_1^2)}\,.
\end{align}
In particular, we impose $k_1^2=-m^2$ to get
\begin{align}
\Delta_{1,0}Q_\text{two-loop}(-m^2)|_{g=0}&=\frac{-1}{72 m^2}\left(\frac{\lambda}{16\pi^2}\right)^2\left(10-\pi^2\right)=\frac{-1}{3}\Pi''_\text{two-loop}(-m^2)|_{g=0}<0\,.
\end{align}
Here, we used eq.~\eqref{eq:two2}. Therefore, we conclude by using eq.~\eqref{eq:wtcoeff2} that the leading order contribution to $R'(0)$ from the quartic coupling $\lambda$ arises at the two-loop level, which can be computed as 
\begin{align}
R'_\text{two-loop}(0)|_{g=0}=\frac{-1}{6}\Pi''_\text{two-loop}(-m^2)|_{g=0}=\frac{-1}{144 m^2}\left(\frac{\lambda}{16\pi^2}\right)^2\left(10-\pi^2\right)<0\,.\label{eq:r2}
\end{align}
We find again that $R'_\text{two-loop}(0)|_{g=0}$ as well as $\Pi''_\text{two-loop}|_{g=0}$ is independent of the renormalization scheme. 
From eqs.~\eqref{eq:r1} and \eqref{eq:r2}, we confirm that $\cgrt$ is given by (11) and (12), and that $\cgrt$ is independent of the renormalization scheme at least within the range of our approximations.

\section{Appendix IV. Computation of $\cgro$}
\label{sec:cgro}
\begin{figure}[tbp]
 \centering
  \includegraphics[width=.8\textwidth, trim=60 200 70 80,clip]{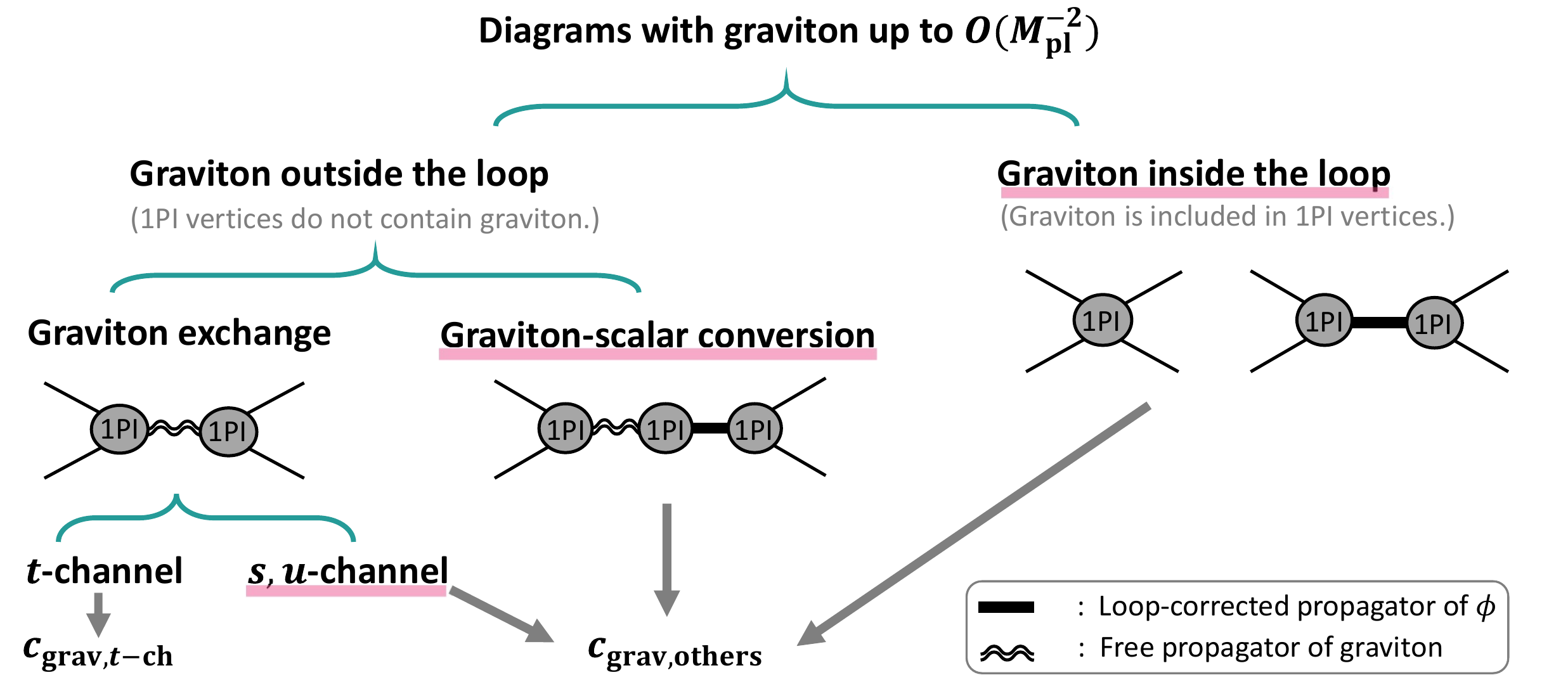}
 \caption{Classification of diagrams with graviton up to $\mathcal{O}(\Mpl^{-2})$. A solid thick line and a thin double wavy line denote a loop-corrected propagator of $\phi$ and a free propagator of graviton, respectively. Diagrams which contribute to $\cgro$ are underlined: (1) $s$, $u$-channel graviton exchange diagrams, (2) diagrams with a graviton-scalar conversion, and (3) those with a graviton propagator inside the loop. The $t$-channel graviton exchange diagrams contribute to $\cgrt$.}
 \label{fig:classification}
\end{figure} 
In this section, we compute $\cgro$ which represents one-loop contributions from (1) the $s,u$-channel graviton exchange diagrams, (2) diagrams with a graviton-scalar conversion, and (3) those with a graviton propagator inside the loop: the classification of diagrams with graviton up to $\mathcal{O}(\Mpl^{-2})$ is shown in FIG.~\ref{fig:classification}. The final goal is to show eq.~(13), which justifies to ignore $\cgro$ when discussing the implication of positivity bound. Throughout the section, we use the harmonic gauge, in which the tree-level graviton propagator of momentum $q$ reads
\begin{align}
\frac{1}{\Mpl^2}\frac{-iP^{(d)}_{\mu\nu\rho\sigma}}{q^2-i\epsilon}
\end{align}
in $d$-dimensions and $P^{(d)}_{\mu\nu\rho\sigma}$ is defined by
\begin{align}
P^{(d)}_{\mu\nu\rho\sigma}\coloneqq \frac{1}{2}\left[\eta_{\mu\rho}\eta_{\nu\sigma}+\eta_{\mu\sigma}\eta_{\nu\rho}-\frac{2}{d-2}\eta_{\mu\nu}\eta_{\rho\sigma}\right]\,.
\end{align}

\begin{figure}[tbp]
 \centering
  \includegraphics[width=.7\textwidth, trim=10 380 250 90,clip]{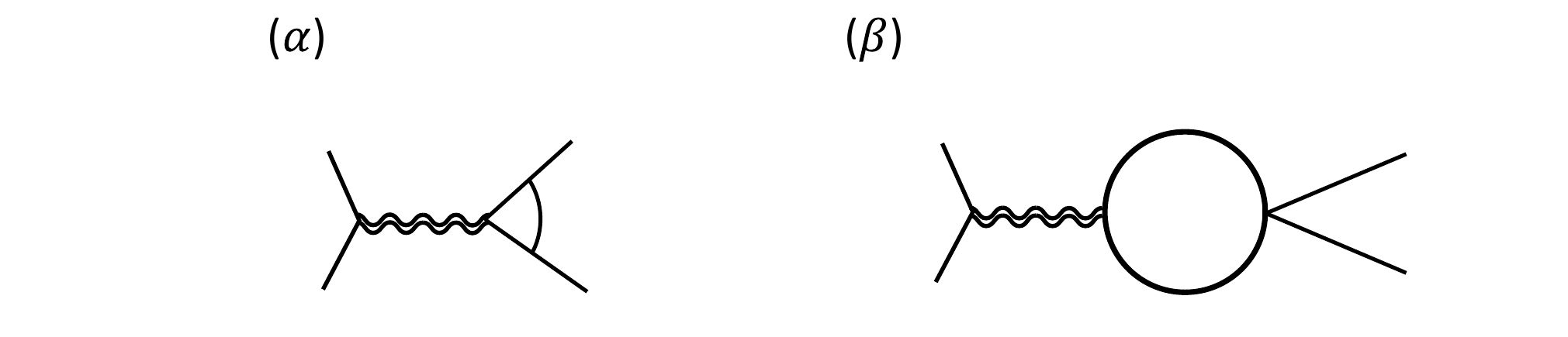}
 \caption{
One-loop graviton exchange diagrams relevant for $\cgr$ at order $\Mpl^{-2}$.
Each type has both $s$- and $u$-channel versions. There are other one-loop diagrams for $s,u$-channel graviton exchange, but they do not contribute to $\cgr$.}
 \label{fig:grav_diag1} 
\end{figure}

\subsection{A. $s,u$-channel graviton exchange}
We start with one-loop corrections to the $s,u$-channel graviton exchange diagrams. The relevant diagrams are shown in  FIG.~\ref{fig:grav_diag1}. Below, we evaluate the contributions from each diagram to $\cgro$, which are denoted as $\cgro|_{(\alpha)}$ and $\cgro|_{(\beta)}$.
\paragraph{Diagram ($\alpha$).} 
We firstly compute the $s$- and $u$-channel pieces of the diagram ($\alpha$) as 
\begin{align}
&\scat_{(\alpha)}(s,t)|_{s,u\text{-}{\rm channel}}\no\\
&=\frac{-2g^2}{\Mpl^2s}V^{\mu\nu}_{\rm tree}(k_1,k_2)P^{(d)}_{\mu\nu\rho\sigma}\tilde\mu^{4-d}\,(-i)\loopint V^{\rho\sigma}_{\rm tree}(-\ell,\ell+k_1+k_2)i\Delta(\ell)i\Delta(\ell+k_1+k_2)i\Delta(\ell-k_4)+(s\leftrightarrow u)\no\\
&=\frac{g^2}{8\pi^2\Mpl^2s}\left\{\frac{s+2m^2}{\varepsilon}-\int_0^1\mathrm{d}x\int^{1-x}_0\mathrm{d}y\,\left[s+(s+2m^2)\ln\left(\frac{\widetilde D_{xy}(s)}{\mu^2}\right)+\frac{A(s,t)}{\widetilde D_{xy}(s)}\right]\right\}+(s\leftrightarrow u)\,, \label{eq:grdig1}
\end{align}
where 
\begin{align}
&\widetilde D_{xy}(s)\coloneqq (x^2-x+1)m^2-y(1-x-y)s\,,\\
&A(s,t)\coloneqq x^2(s+t)t-(x^2s+4x^2t+s)m^2+(4x^2-2)m^4\,.
\end{align}
The UV divergence can be renormalized by $\delta Z_m$ after other diagrams which are omitted in FIG.~\ref{fig:grav_diag1} are also taken into account. Eq.~\eqref{eq:grdig1} shows that $\scat_{(\alpha)}(s,0)|_{s,u\text{-}{\rm channel}}$ satisfies the Froissart bound and the standard analyticity properties in the complex $s$-plane. We can then compute the contribution from the diagram ($\alpha$) to $\cgro$ as
\begin{align}
\cgro|_{(\alpha)}=\frac{4}{\pi}\int^\infty_{\lth^2}\mathrm{d}s'\,\frac{\im\,\scat_{(\alpha)}(s',0)}{(s'-2m^2)^3}\simeq\frac{g^2}{8\pi^2\Mpl^2\lth^4}\,,\label{eq:gro1}
\end{align}
analogously to eq.~\eqref{eq:ngdis}. Here, we used 
\begin{align}
\im\,\scat_{(\alpha)}(s,0)|_{s\geq4m^2}=\frac{g^2}{8\pi\Mpl^2}\left[\sqrt{\frac{s-4m^2}{s}}\frac{s-2m^2}{2s}+\frac{1}{\sqrt{s(s-4m^2)}}\frac{m^2(s+m^2)}{s}\ln\,\biggl(\frac{s-3m^2}{m^2}\biggr)\right]\,.
\end{align}
Note that the $t$-channel version of the diagram ($\alpha$) contributes to $\cgrt$, giving rise to the first term of (11). From \eqref{eq:gro1}, it is found that $\cgro|_{(\alpha)}$ is much smaller than the first term of $\cgrt$ when $\lth^2\gg m^2$. Practically, the contributions from $s,u$-channel diagrams are negligible when $\lth\gtrsim10m$. 

\paragraph{Diagram ($\beta$).}
We compute the $s$- and $u$-channel pieces of the diagram ($\beta$) as 
\begin{align}
&\scat_{(\beta)}(s,t)|_{s,u\text{-}{\rm channel}}\no\\
&=\frac{\lambda}{\Mpl^2s}V^{\mu\nu}_{\rm tree}(k_1,k_2)P^{(d)}_{\mu\nu\rho\sigma}\tilde\mu^{4-d}(-i)\loopint V^{\rho\sigma}_{\rm tree}(-\ell,\ell+k_1+k_2)i\Delta(\ell)i\Delta(\ell+k_1+k_2)+(s\leftrightarrow u)\no\\
&=-\frac{5\lambda m^2}{24\pi^2\Mpl^2\varepsilon}-\frac{\lambda m^2}{16\pi^2\Mpl^2}+\frac{\lambda}{16\pi^2\Mpl^2}\int^1_0\mathrm{d}x\,(x-x^2)\left[(s+2m^2)\ln\left(\frac{D_x(s)}{\mu^2}\right)+(s\leftrightarrow u)\right]\,.\label{eq:eta1}
\end{align}
We can renormalize this UV divergence by adding an $\mathcal{O}(\lambda/\varepsilon)$ term to $\delta Z_{R\phi^2}$ as explained around eq.~\eqref{eq:defd1}. It is now obvious that $\cgro|_{(\beta)}$ can be computed as 
\begin{align}
\cgro|_{(\beta)}&=\frac{4}{\pi}\int^\infty_{\lth^2}\mathrm{d}s'\,\frac{\im\,\scat_{(\beta)}(s',0)}{\left(s'-2m^2\right)^3}\simeq\frac{-\lambda}{24\pi^2\Mpl^2\lth^2}\,,
\end{align}
where we used
\begin{align}
&\left.\im\,\scat_{(\beta)}(s,0)\right|_{s\geq 4m^2}=-\frac{\lambda}{96\pi\Mpl^2}\sqrt{\frac{s-4m^2}{s}}\frac{(s+2m^2)^2}{s}\,.
\end{align}
We then have 
\begin{align}
\cgro|_{(\alpha)}+\cgro|_{(\beta)}\simeq \frac{g^2}{8\pi^2\Mpl^2\lth^4}-\frac{\lambda}{24\pi^2\Mpl^2\lth^2}\,,
\end{align}
confirming the estimation (13). 

\subsection{B. Graviton-scalar conversion}
\begin{figure}[tbp]
 \centering
  \includegraphics[width=.45\textwidth, trim=10 380 500 90,clip]{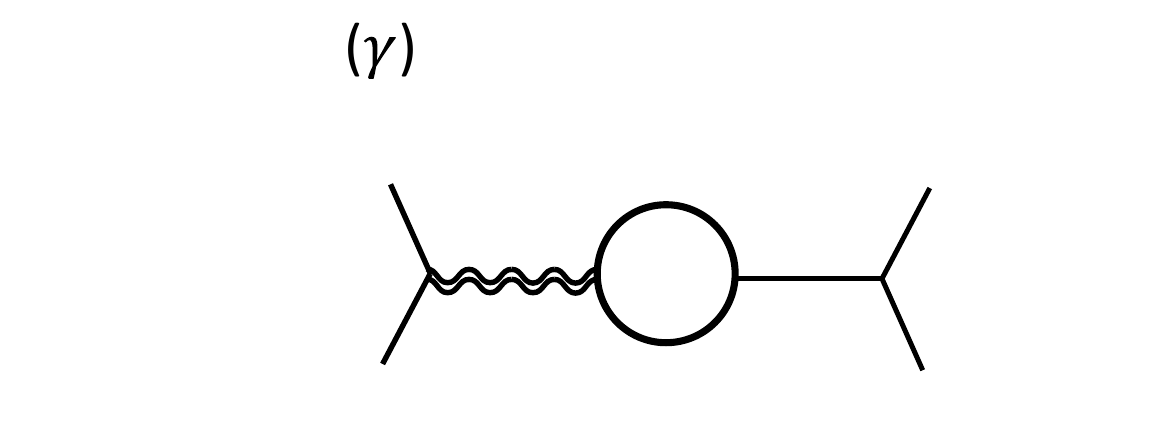}
 \caption{The one-loop diagram with a graviton-scalar conversion that are relevant for $\cgr$ at order $\Mpl^{-2}$. For this type, the $s,u$-channel diagrams have a nonzero contribution to $\cgr$, but the $t$-channel diagram does not. Also, there are other one-loop diagrams with a graviton-scalar conversion, but they do not contribute to $\cgr$.}
\label{fig:grav_diag1-2} 
\end{figure}

We then consider one-loop diagrams with a graviton-scalar conversion. The relevant diagrams are shown in  FIG.~\ref{fig:grav_diag1-2}, whose contribution to $\cgro$ is denoted as $\cgro|_{(\gamma)}$. Note that only the $s$- and $u$-channel diagrams have nonzero contributions to $\cgro$. These diagrams are computed as 
\begin{align}
&\scat_{(\gamma)}(s,t)|_{s,u\text{-}{\rm channel}}\no\\
&=\frac{-g^2}{\Mpl^2s(m^2-s)}V^{\mu\nu}_{\rm tree}(k_1,k_2)P^{(d)}_{\mu\nu\rho\sigma}\tilde\mu^{4-d}(-i)\loopint V^{\rho\sigma}_{\rm tree}(-\ell, \ell+k_1+k_2)i\Delta(\ell)i\Delta(\ell+k_1+k_2)+(s\leftrightarrow u)\no\\
&=\frac{-g^2}{16\pi^2\Mpl^2(m^2-s)}\left[\frac{-(s+2m^2)}{3\varepsilon}+\int^1_0\mathrm{d}x\,(x-x^2)\left((s+2m^2)\ln\left(\frac{D_x(s)}{\mu^2}\right)-2m^2\right)\right]+(s\leftrightarrow u)\,.
\end{align}
UV divergences can be renormalized by adding an $\mathcal{O}(g^2/\varepsilon)$ term to $\delta Z_{R\phi}$. This expression shows that  $\scat_{(\gamma)}(s,0)|_{s,u\text{-}{\rm channel}}$ satisfies the Froissart bound and the standard analyticity properties in the complex $s$-plane. We can then compute $\cgro|_{(\gamma)}$ as
\begin{align}
\cgro|_{(\gamma)}=\frac{4}{\pi}\int^\infty_{\lth^2}\mathrm{d}s'\,\frac{\im\,\scat_{(\gamma)}(s',0)}{(s'-2m^2)^3}\simeq-\frac{g^2}{48\pi^2\Mpl^2\lth^4}\,,\label{eq:gro2}
\end{align}
confirming the estimation (13). 
Here, we used  
\begin{align}
\im\,\scat_{(\gamma)}(s,0)|_{s\geq 4m^2}=\frac{g^2}{96\pi\Mpl^2}\frac{s-4m^2}{s}\frac{\left(s+2m^2\right)^2}{s(m^2-s)}\,.
\end{align}

\subsection{C. Graviton inside loop}

Finally, we evaluate the contributions from diagrams with a graviton inside the loop. All the relevant diagrams are shown in FIG.~\ref{fig:grav_loop3} in which all the possible assignment of external momenta should be considered. As we shall see below, some of these diagrams suffer from infrared (IR) divergences. It will be necessary to consider the dressed amplitude appropriately to resolve this issue, precisely speaking. In the present analysis, however, we simply introduce the fictitious graviton mass $m_g$ to deform the free graviton propagator with momentum $q$ in $d$-dimensions as 
\begin{align}
\frac{1}{\Mpl^2}\frac{-i\,P^{(d)}_{\mu\nu\rho\sigma}}{q^2-i\epsilon}\to \frac{1}{\Mpl^2}\frac{-i\,P^{(d)}_{\mu\nu\rho\sigma}}{q^2+m_g^2-i\epsilon}\,,
\end{align}
to regulate the IR divergences. We then compute $\cgro$ to verify the order estimation (13). We suppose that this prescription is enough for the order-of-magnitude estimate of $\cgro$. For later convenience, we introduce the notation $i\Delta_g(k)\coloneqq 1/(k^2+m_g^2-i\epsilon)$.

To compute $\cgro$, we use the fact that all the diagrams (A), (B), and (C) give the analytic amplitudes which behave mildly at large $|s|$ to satisfy the relations
\begin{align}
\cgro|_{\rm FIG.\ref{fig:grav_loop3}}=\frac{4}{\pi}\int^\infty_{\lth^2}\mathrm{d}s'\,\frac{\im\,\scat(s',0)|_{\rm FIG.\ref{fig:grav_loop3}}}{\left(s'-2m^2\right)^3}\,.\label{useful}
\end{align}
Firstly, we shall check the high-energy behavior of $\scat(s,0)|_{\rm FIG.\ref{fig:grav_loop3}}$. Then, we compute the imaginary part to obtain $\cgro$.

\begin{figure}[tbp]
 \centering
  \includegraphics[width=.8\textwidth, trim=75 230 90 60,clip]{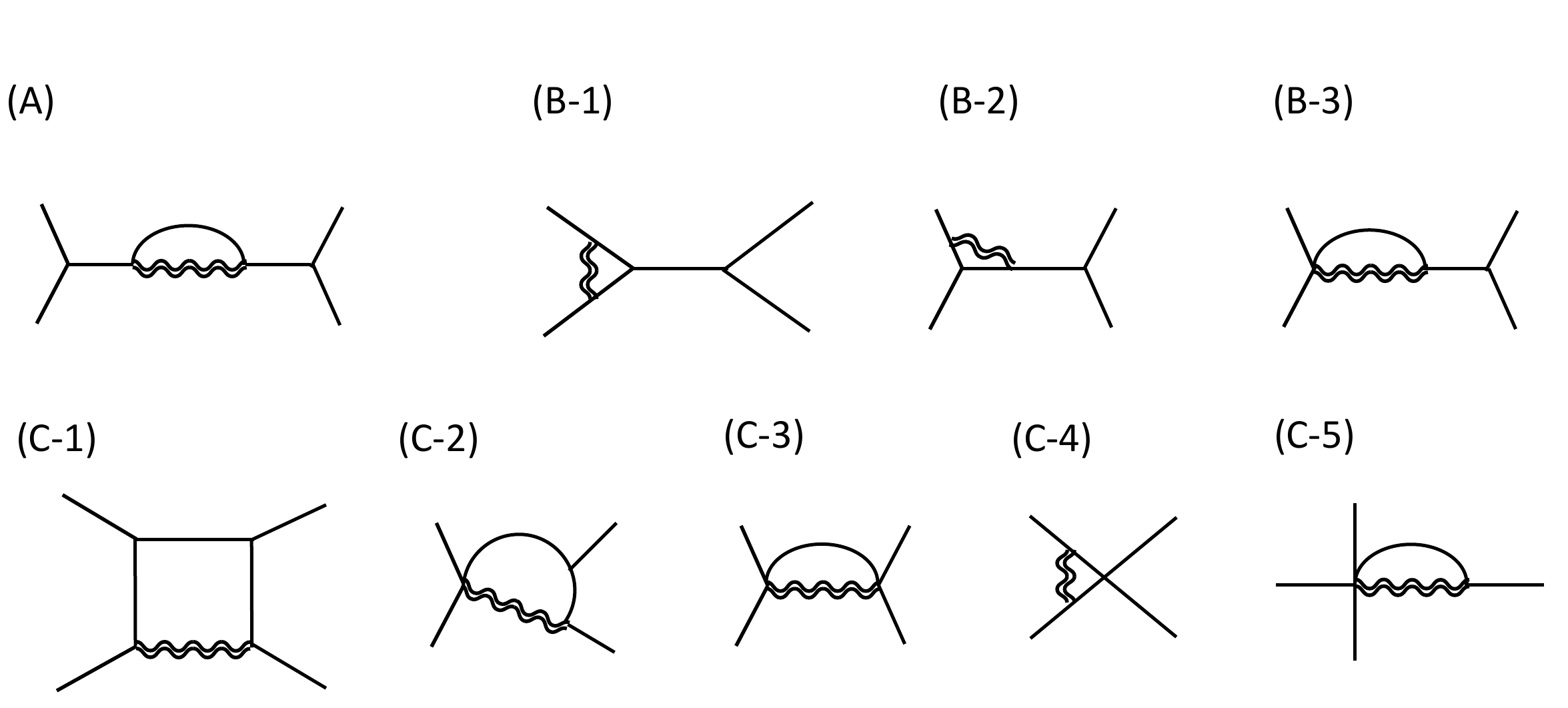}
 \caption{Diagrams with a graviton propagator inside loops which are of $\mathcal{O}(\Mpl^{-2})$. Again, all the possible assignment of external momenta have to be considered. The diagrams (A), (B), and (C) can be understood as the contributions from the $\mathcal{O}(\Mpl^{-2})$ corrections to the 1PI self-energy, the effective $\phi^3$ vertex, and the effective $\phi^4$ vertex, respectively.}
 \label{fig:grav_loop3} 
\end{figure}

\subsubsection{1. $s^2$-boundedness}
We investigate the high-energy behavior of $\scat|_{\rm FIG.\ref{fig:grav_loop3}}$. We used ``package-X''~\cite{Patel:2015tea} to perform the loop integrals  and obtain the asymptotic behavior of the amplitudes in the Regge limit. 

\paragraph{Diagram (A).}
The diagram (A) is the contribution from the $\mathcal{O}(\Mpl^{-2})$ correction to 1PI self-energy of $\phi$ to $\scat$.  
This diagram is IR finite even in the limit $m_g\to0$ and its asymptotic behavior in the Regge limit is 
\begin{align}
\scat_{\rm (A)}(s,0)&\sim \frac{g^2}{\Mpl^2}\frac{1}{(m^2-s)^2}\tilde\mu^{4-d}\loopint\,V^{\mu\nu}_{\rm tree}(\ell,k_1+k_2)P^{(d)}_{\mu\nu\rho\sigma}V^{\rho\sigma}_{\rm tree}(\ell,k_1+k_2)i\Delta(\ell)i\Delta_g(\ell+k_1+k_2)\no\\
&\quad+(s\leftrightarrow t)+(s\leftrightarrow u)\sim \mathcal{O}(s^{0})\,.\label{grloopA}
\end{align}
This behavior satisfies the Froissart bound.

\paragraph{Diagrams (B).}
The diagrams (B) are the contributions from the $\mathcal{O}(\Mpl^{-2})$ correction to 1PI effective $\phi^3$ vertex to $\scat$. The diagrams (B-1) and (B-2) are IR divergent, while the diagram (B-3) is IR finite. The asymptotic behavior of them in the Regge limit can be estimated as 
\begin{subequations}
\label{grloopB}
\begin{align}
\scat_{\rm (B\text{-}1)}(s,0)&\sim \frac{g^2}{\Mpl^2}\frac{1}{m^2-s}\tilde\mu^{4-d}\loopint\,V^{\mu\nu}_{\rm tree}(\ell,k_2)P^{(d)}_{\mu\nu\rho\sigma}V^{\rho\sigma}_{\rm tree}(\ell+k_1+k_2,-k_1)i\Delta(\ell)i\Delta(\ell+k_1+k_2)i\Delta_g(\ell+k_2)\no\\
&\quad+(s\leftrightarrow t)+(s\leftrightarrow u)\sim \mathcal{O}(\ln(s))\,,\\
\scat_{\rm (B\text{-}2)}(s,0)&\sim \frac{g^2}{\Mpl^2}\frac{1}{m^2-s}\tilde\mu^{4-d}\loopint\,V^{\mu\nu}_{\rm tree}(\ell,k_1+k_2)P^{(d)}_{\mu\nu\rho\sigma}V^{\rho\sigma}_{\rm tree}(\ell+k_2,k_1)i\Delta(\ell)i\Delta(\ell+k_2)i\Delta_g(\ell+k_1+k_2)\no\\
&\quad+(s\leftrightarrow t)+(s\leftrightarrow u)\sim \mathcal{O}(s^0)\,,\\
\scat_{\rm (B\text{-}3)}(s,0)&\sim \frac{g^2}{\Mpl^2}\frac{1}{m^2-s}\tilde\mu^{4-d}\loopint\,V^{\mu\nu}_{\rm tree}(\ell,k_1+k_2)P^{(d)}_{\mu\nu\rho\sigma}\eta^{\rho\sigma}i\Delta(\ell)i\Delta_g(\ell+k_1+k_2)+(s\leftrightarrow t)+(s\leftrightarrow u)\no\\
&\sim\mathcal{O}(\ln(s))\,.
\end{align}
\end{subequations}
We conclude that the diagrams (B) are consistent with the Froissart bound.

\paragraph{Diagrams (C).}
The diagrams (C) are contributions from the $\mathcal{O}(\Mpl^{-2})$ corrections to the 1PI effective $\phi^4$ vertex to $\scat$. The diagrams (C-3) and (C-5) are IR finite, while other diagrams are IR divergent. They behave in the Regge limit asymptotically as
\begin{subequations}
\label{grloopC}
\begin{align}
\scat_\text{(C-1)}(s,0)&\sim\frac{g^2}{\Mpl^2}\tilde\mu^{4-d}\loopint\,\Biggl\{\biggl[i\Delta(\ell)i\Delta(\ell+k_1+k_2)i\Delta(\ell-k_3)i\Delta_g(\ell+k_1)\no\\
&\quad\times  V^{\mu\nu}_{\rm tree}(\ell,k_1)P^{(d)}_{\mu\nu\rho\sigma}V^{\rho\sigma}_{\rm tree}(\ell+k_1+k_2,-k_2)+(k_1\leftrightarrow k_3)\biggr]+(k_3\leftrightarrow k_4)\Biggr\}+(s\leftrightarrow t)+(s\leftrightarrow u)\sim\mathcal{O}(s)\,,\\
\scat_\text{(C-2)}(s,0)&\sim \frac{g^2}{\Mpl^2}\tilde\mu^{4-d}\loopint\,\left[V^{\mu\nu}_{\rm tree}(\ell,k_1)P^{(d)}_{\mu\nu\rho\sigma}\eta^{\rho\sigma}i\Delta(\ell)i\Delta(\ell-k_2)i\Delta_g(\ell+k_1)+(k_1\leftrightarrow k_2)\right]+(s\leftrightarrow t)+(s\leftrightarrow u)\no\\
&\sim\mathcal{O}(\ln(s))\,,\\
\scat_\text{(C-3)}(s,0)&\sim \frac{g^2}{\Mpl^2}\tilde\mu^{4-d}\loopint\,\eta^{\mu\nu}P^{(d)}_{\mu\nu\rho\sigma}\eta^{\rho\sigma}i\Delta(\ell)i\Delta_g(\ell+k_1+k_2)+(s\leftrightarrow t)+(s\leftrightarrow u)\sim\mathcal{O}(\ln(s))\,,\\
\scat_\text{(C-4)}(s,0)&\sim \frac{\lambda}{\Mpl^2}\tilde\mu^{4-d}\loopint\,V^{\mu\nu}_{\rm tree}(\ell,k_2)P^{(d)}_{\mu\nu\rho\sigma}V^{\rho\sigma}_{\rm tree}(\ell+k_1+k_2,-k_1)i\Delta(\ell)i\Delta(\ell+k_1+k_2)i\Delta_g(\ell+k_2)\no\\
&\hspace{3cm}+(s\leftrightarrow t)+(s\leftrightarrow u)\sim\mathcal{O}(s\ln(s))\,,\\
\scat_\text{(C-5)}(s,0)&\sim\frac{\lambda}{\Mpl^2} \sum_{j=1}^4\tilde\mu^{4-d}\loopint\,V^{\mu\nu}_{\rm tree}(\ell,k_j)P^{(d)}_{\mu\nu\rho\sigma}\eta^{\rho\sigma}i\Delta(\ell)i\Delta_g(\ell+k_j)\sim\mathcal{O}(s^0)\,,
\end{align}
\end{subequations}
consistently with the Froissart bound. We confirm the mild behavior of $\scat(s,0)|_{\rm FIG.\ref{fig:grav_loop3}}$, leading to \eqref{useful} together with the analyticity.

\subsubsection{2. Imaginary part}
The imaginary part of each diagram in the forward limit can be computed by using the optical theorem as 
\begin{align}
&\im\,\scat_{\rm (A)}(s,0)|_{s\gg m^2}\simeq\frac{g^2m^2}{4\pi\Mpl^2s}\,,\qquad \im\,\scat_{\rm (B\text{-}1)}(s,0)|_{s\gg m^2}\simeq \frac{-g^2}{16\pi\Mpl^2}\left[\ln\left(\frac{s}{m_g^2}\right)-1\right]\,,\\
&\im\,\scat_{\rm (B\text{-}2)}|_{s\gg m^2}\simeq\frac{g^2}{8\pi\Mpl^2}\,,\quad \im\,\scat_{\rm (B\text{-}3)}(s,0)|_{s\gg m^2}\simeq\frac{g^2}{8\pi\Mpl^2}\,,\quad \im\,\scat_{\rm (C\text{-}1)}(s,0)|_{s\gg m^2}\simeq\frac{g^2s}{16\pi\Mpl^2m^2}\ln\left(\frac{m^2}{m_g^2}\right)\,,\\
&\im\,\scat_{\rm (C\text{-}2)}(s,0)|_{s\gg m^2}\simeq\frac{g^2}{4\pi\Mpl^2}\,,\quad \im\,\scat_{\rm (C\text{-}3)}(s,0)|_{s\gg m^2}\simeq-\frac{g^2}{4\pi\Mpl^2}\,,\\
&\im\,\scat_{\rm (C\text{-}4)}(s,0)|_{s\gg m^2}\simeq-\frac{\lambda s}{16\pi\Mpl^2}\left[\ln\left(\frac{s}{m_g^2}\right)-1\right]\,,\quad \im\,\scat_{\rm (C\text{-}5)}(s,0)=0\,,
\end{align}
where higher order terms suppressed by some positive powers of $(m^2/s)$ are omitted. This shows that the imaginary part of diagrams shown in FIG.~\ref{fig:grav_loop3} are dominated by the diagram (C-1) and (C-4), resulting in  
\begin{align}
\cgro|_{\rm FIG.\ref{fig:grav_loop3}}
\simeq\frac{1}{4\pi^2\Mpl^2\lth^2}\left[\frac{g^2}{m^2}\ln\left(\frac{m^2}{m_g^2}\right)-\lambda\ln\left(\frac{\lth^2}{m_g^2}\right)\right]\sim\mathcal{O}\left(\frac{(g/m)^2}{\Mpl^2\lth^2},\,\frac{\lambda}{\Mpl^2\lth^2}\right)\,,\label{eq:est2}
\end{align}
confirming the estimation (13).
\end{widetext}

%\bibliography{positivity_2020_2}

\end{document}